\documentclass{jfm}

\usepackage{graphicx}
\usepackage{newtxtext}
\usepackage{newtxmath}
\usepackage{natbib}
\usepackage[justification=justified,format=plain]{caption}
\usepackage{hyperref}
\hypersetup{
    colorlinks = true,
    urlcolor   = blue,
    citecolor  = black,
}

\newcommand{\RomanNumeralCaps}[1]
\linenumbers

\title{Quantifying the liquid flow between a soap film and a vertical meniscus}

\author{Alexandre Vigna-Brummer\aff{1}, 
Simon Cox\aff{2},
Médéric Argentina\aff{1}, 
Christophe~Brouzet\aff{1}, 
and Christophe Raufaste\aff{1,3}\corresp{\email{christophe.raufaste@univ-cotedazur.fr}}}

\affiliation{\aff{1}Universit\'e C\^ote d'Azur, CNRS, INPHYNI, France
\aff{2}Department of Mathematics, Aberystwyth University, Ceredigion SY23 3BZ, UK
\aff{3}Institut Universitaire de France (IUF), France}

\begin{document}
\maketitle

\begin{abstract}
Fluid exchange between a soap film and its bounding menisci governs film drainage and stability, with direct implications for the lifetime of surface bubbles and liquid foams. Despite recent advances, a quantitative characterization of this coupling, associated with the phenomenon of marginal regeneration, remains incomplete. The volumetric flux per unit length of contact follows a well-established scaling law involving geometrical parameters such as the film height and the meniscus radius of curvature. However, the dimensionless prefactor of this relation—the flux coefficient—remains difficult to determine for vertical menisci because of the complex and intermittent flows occurring at the film–meniscus interface. 
Here, we quantify this flux into the meniscus generated by inserting a solid plate into a vertical soap film. 
We consider both vertical and inclined plates and further investigate the effects of plate inclination, height, and width. Focusing on the  dynamics of the growth of the meniscus driven by liquid supplied by the film, we analyze both steady and transient regimes resulting from the interplay between capillary pressure, gravity, and liquid exchange. Combining experiments, numerical simulations, and theoretical modelling, we determine the flux coefficient using three independent methods and show that it remains constant over the range of parameters explored.
\end{abstract}

\maketitle


\section{Introduction} 

Soap films have long fascinated scientists due to their iridescent color patterns, which arise from the interaction of light with these liquid membranes having micron or submicron thicknesses~\citep{gochev2016chronicles,ziapkoff_white_2026}. While single soap films suspended on a frame are primarily academic model systems for studying surface-tension–driven phenomena~\citep{lovett1994, weaire2001, cantat2013foams}, their stability is of paramount importance in many natural and industrial contexts. In particular, soap-film stability plays a key role in the lifetime of surface bubbles and the generation of sea spray aerosols \citep{woodcock_giant_1953, macintyre_flow_1972, wu_evidence_1981, leifer_secondary_2000, bird_daughter_2010, deike_mass_2022}. In bubble assemblies or liquid foams, soap films separate gas bubbles, and their stability is crucial for applications across the food, cosmetics, and pharmaceutical industries, as well as in energy and environmental technologies \citep{stevenson_foam_2012, sun_research_2023, chaudhry_recent_2024, harshini_innovative_2024}.

However, these structures are inherently transient and their thickness evolves with time, ultimately leading to rupture. Soap films are connected to solid frames, liquid baths (in the case of surface bubbles), or other bubbles in liquid foams—where they are referred to as foam films—through menisci known as Plateau borders \citep{cantat2013foams}.  The concave shape of Plateau borders, characterized by a radius of curvature $r$, generates a capillary suction due to a pressure $-\gamma/r$ relative to the atmosphere, where $\gamma$ is the surface tension of the liquid-air interface. This capillary pressure induces a net volumetric flux of liquid from the film into the surrounding meniscus~\citep{cantat_drainage_2026}, contributing to film thinning in horizontal films \citep{sheludko_thin_1967, sonin_role_1993, joye_asymmetric_1994, velev_investigation_1995, cascao_pereira_bike-wheel_2001, yaminsky_stability_2010, chatzigiannakis_thin_2021, andrieux_microfluidic_2021} and surface bubbles \citep{lhuissier2012bursting, frostad_dynamic_2016, bhamla_placing_2016, bhamla2017interfacial, lin_influence_2018, liu_dynamic_2018, miguet2021marginal}, 
while the exact mechanism remains an open question for vertical soap and foam films \citep{mysels1959soap, hudales_marginal_1990, berg_experimental_2005, elias_magnetic_2005, tan_thinning_2010, sett_gravitational_2013, seiwert_velocity_2017, yu_stability_2022, monier_self-similar_2024, auregan_drainage_2025}.

The liquid exchange between the film and meniscus is not uniform but is influenced by the instability known as marginal regeneration, a mechanism first proposed by \citet{mysels1959soap} and extended by \citet{gros2021marginal}. During this process, the initial film of thickness $h$ feeds into the meniscus, while thinner patches—thin film elements (TFEs) with thickness $h_\text{TFE} < h$—emerge from the meniscus, resulting in a net mass loss from the film, described in more detail in the Appendix. Quantitatively, the suction induces a flux per unit length of contact $q$ from the film into the meniscus. Writing $\eta$ for the dynamic viscosity of the liquid, the mass balance combined with the lubrication equation at the contact line between the film and the meniscus leads to the scaling \citep{mysels1959soap, lhuissier2012bursting, gros2021marginal, cantat_drainage_2026}:
\begin{equation}\label{Eq:linear_flux} 
    q = k_e\frac{\gamma}{\eta}\frac{h^{5/2}}{r^{3/2}}, 
\end{equation} 
as verified experimentally \citep{vigna-brummer_flowing_2025, monier_Centrifugal_2025}, where $k_e$ is a dimensionless flux coefficient, whose determination is the main objective of the present study.

For a model horizontal meniscus, $k_e$ depends on the fraction $\xi$ of the contact line between meniscus and film covered by TFEs. Then $k_e$ vanishes in the limiting cases $\xi = 0$ and $\xi = 1$, and reaches a maximum value of approximately 3.6$\times10^{-2}$ for $\xi \simeq 0.87$ (see Appendix for a figure and the analytical expression derived from \citet{gros2021marginal} and \citet{cantat_drainage_2026}). In experiments on horizontal films reported in~\citet{gros2021marginal}, TFEs both grow and merge over time, leading to an evolution of $\xi$ from about 0.8 to 0.94–0.97. This corresponds to expected values of $k_e$ in the range 2.3$\times10^{-2}$ to 3.6$\times10^{-2}$ during the course of those experiments.

The scaling proposed in Eq.~\eqref{Eq:linear_flux} accounts for the thickness evolution of surface bubbles  \citep{lhuissier2012bursting, miguet2021marginal}. In that case, TFEs detach from the horizontal meniscus and migrate upward due to buoyancy \citep{adami2014capillary, plateau1873, couder1989hydrodynamics}. The same scaling is expected to apply at the bottom of vertical films where TFEs can be characterized quantitatively \citep{nierstrasz1998marginal, yi_quantitative_2025}. The situation becomes more complex along the vertical menisci of vertical films, which are known to dominate fluid exchange with soap films \citep{mysels1959soap, hudales1990marginal, berg_experimental_2005, seiwert2017velocity, monier_self-similar_2024, auregan_drainage_2025}. Along these vertical menisci, TFEs migrate very rapidly, leading to intermittent and complex flows composed of irregular TFEs. Although the same scaling still holds \citep{vigna-brummer_flowing_2025}, this behavior makes the exchange process difficult to analyze directly and complicates the estimation of an effective value of $k_e$, even under steady-state conditions.

In this article, building upon previous work \citep{vigna-brummer_flowing_2025}, we investigate the flux exchange and aim to measure the dimensionless flux coefficient $k_e$ for the contact between a vertical film and a vertical meniscus. We also examine the effect of inclination by varying the meniscus angle 
with respect to the vertical.  
Our approach combines experimental measurements with numerical simulations. Specifically, we insert solid plates into a vertical soap film to generate menisci. By systematically varying the plate geometry (height, width, inclination) and the film thickness, we first analyze the meniscus profile in steady state. Measurements of the shape of the meniscus, coupled with an analysis of the liquid drainage dynamics within it, allow us to quantify the flux and evaluate the flux coefficient $k_e$. Second, using numerical simulations to link the meniscus shape and volume, we examine the growth of the meniscus to obtain a direct measurement of the coefficient in the transient regime. Finally, we discuss our findings in the context of the existing literature.

\section{Materials and Methods}
\subsection{Experimental set-up\label{sec:experimentalsetup}}

\begin{figure*}
    \includegraphics[height=0.38\linewidth]{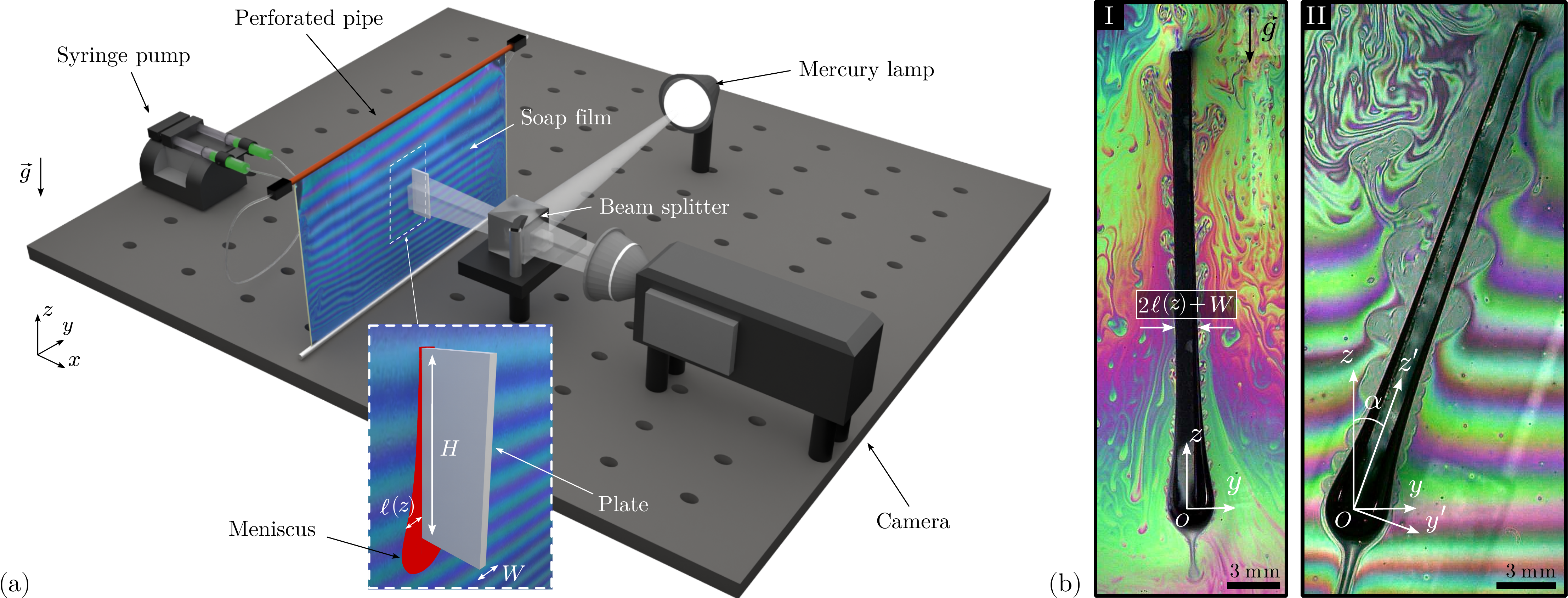}
    \caption{(a)~Experimental set-up of a vertical soap film with controlled thickness profile. Inset: Plate  of height~$H$ and width~$W$ introduced into the vertical soap film. The meniscus is drawn around the plate, highlighted in red.
    (b)~Typical snapshots of a glass plate ($H=26$~mm and $W=1$~mm) in a vertical soap film. (I)~Vertical plate ($\alpha=0^{\circ}$) and $h\simeq 0.5 \ \mu$m. (II)~Plate inclined at~$\alpha=20^{\circ}$ with $h\simeq3.5 \ \mu$m. Note that the magnification is slightly different between the two images. The corresponding movies are available in the Supplementary Material (movies 1 and 2, respectively).}
    \label{Fig:setup_profile}
\end{figure*}

Vertical soap films were created within a rectangular frame measuring $150$~mm in height and $250$~mm in width. The frame consists of a horizontal perforated pipe at the top, connected to a horizontal rod at the bottom by two vertical strings, as detailed in Fig.~\ref{Fig:setup_profile}(a). Surfactant solution is injected into the perforated pipe to supply the film and hence counteract gravitational drainage, leading to a stationary thickness profile~\citep{adami2014capillary, vigna-brummer_flowing_2025}, in which the films are thinner at the top than at the bottom. A constant flow rate between $0.01$ and $0.02$~cm$^{3}$.s$^{-1}$
was maintained using a syringe pump. Two different surfactant solutions were tested to confirm the universality of the results~\citep{vigna-brummer_flowing_2025}: a 10 wt \% solution of a commercial mixture (Dreft) and a 0.1 wt \% solution of 2/3 SLES and 1/3 CAPB. Both solutions were prepared with deionized water. Varying the type of surfactant modifies the mean thickness of the soap film. 
Both solutions lead to mobile films, characterized by in-plane recirculation of film elements during drainage \citep{tregouet2021instability}.

Plates, of width~$W$ and height~$H$, were introduced into the vertical soap film, as shown in Fig.~\ref{Fig:setup_profile}(a). They are held in a fixed position protruding by a few millimeters to prevent the film from pinning to the leading edge of the plate. Experiments were performed with both vertical and inclined plates, where the angle~$\alpha$ denotes the orientation of the plate with respect to the vertical: $\alpha=0$ corresponds to the vertical position and $90^\circ$ to the horizontal. The origin of the coordinate systems is taken at the bottom of the plate and in the middle of the plate within its lateral dimension, as illustrated in Fig.~\ref{Fig:setup_profile}(b). The coordinate system $(O,x,y,z)$ is fixed within the lab frame, while the coordinate system $(O,x',y',z')$ remains attached to the plate, and is rotated with an angle~$\alpha$ when the plate is inclined. The vertical axis~$z$ points upward. The vertical coordinate~$z$ is related to the coordinate~$z'$ along the plate through $z=z'\cos \alpha$. The vertical position of the top of the plate is therefore given by $z= H \cos\alpha$. The right-hand side of the plate is located at $y = W/2$. The transverse coordinate is denoted by $x$ or $x'$.

Two types of materials were used or the plates --  glass and steel -- both of which are wetted by the surfactant solution. 
Plate width, height and inclination are varied within large ranges: $W= 0.05-9.90$~mm,~$H=4-76$~mm, and $\alpha=0^\circ-60^\circ$. 
Once the plate is introduced into the soap film, the meniscus fills up until it becomes unstable at its lower end. At this stage, the meniscus shape becomes stationary and liquid leaks from the bottom in the form of droplets or a jet, as shown in Fig.~\ref{Fig:setup_profile}(b). 
This highlights the liquid exchange between the meniscus and the film, as already observed for rings placed in vertical soap films~\citep{vigna-brummer_flowing_2025}.

The thickness~$h$ of the soap film near the top of the plate is measured using a spectrometer. This parameter can be varied in a large range between $0.5$ and $6.1~\mu$m which can be achieved by combining different effects: (i)~by changing the surfactant solution,  (ii)~by adjusting the plate's vertical position within the film (which is slightly stratified in thickness), (iii)~by stopping the syringe pump and allowing the film to slowly drain under gravity (to obtain the thinnest films).

Typical images such as those shown in Fig.~\ref{Fig:setup_profile}(b) are captured using the optical set-up sketched in Fig.~\ref{Fig:setup_profile}(a). Diffuse light from a mercury lamp is directed onto a large beam splitter, which sends it onto the soap film. The camera then collects the light reflected from the soap film through the beam splitter. 
The meniscus is visible at the contact between the plate and the soap film, drawn in red in Fig.~\ref{Fig:setup_profile}(a) or visible in black (together with the plate) in Fig.~\ref{Fig:setup_profile}(b).
Focusing on the right-hand side of the plate, we denote by $y(z)$ the distance between the mid-plane of the plate and the outer edge of the meniscus, identified as the base of the rising TFEs. In practice, we measure the distance between the positions of the TFEs rising on both sides of the plate and take half of this value.  
For $z>0$, this quantity allows us to define the extent of the meniscus as $\ell(z) = y(z) - W/2$. The radius of curvature $r$ is then deduced by assuming that the film perfectly wets the plate, with a zero contact angle. This assumption leads to the meniscus shape in the $x$–$y$ plane illustrated in Fig.~\ref{Fig:simulation}(a), and to the simple relation $r=\ell$, provided that $r$ is smaller than the characteristic length scale over which $\ell$ varies along~$z$. In addition, the cross-sectional area $A$ of the meniscus can be computed as $A = (2-\pi/2)r^2$. The meniscus is narrowest at the top of the plate, where it has a finite radius of curvature $r(H)=r_H$, and widens toward the bottom of the plate. We therefore measure the meniscus profile along the plate, i.e. $r(z)$. 
Below the plate ($z<0$), $y(z)$ changes abruptly and ultimately reaches zero within a few millimeters. As a consequence, the extent and radius of curvature of the meniscus cannot be unambiguously defined in this region. Nevertheless, for the sake of continuity when representing both experimental and numerical data, we continue to use the relation $r(z) = y(z) - W/2$ below $z=0$, while keeping in mind that this quantity no longer corresponds to a true radius of curvature. In this region, $r$ may even take negative values as $z$ approaches the lowest part of the meniscus.

Regarding the accuracy of the profile measurements, there is a trade-off between capturing the full plate field of view and achieving high spatial resolution; ensuring that the whole of the plate is visible results in larger error bars for longer plates when measuring $\ell(z)$. Note that in experiments where only the radius of curvature $r_H$ at the top of the plate is measured, the camera is zoomed in on the upper region to reduce measurement errors.

\subsection{Numerical set-up}

\begin{figure*}
    \includegraphics[width=\linewidth]{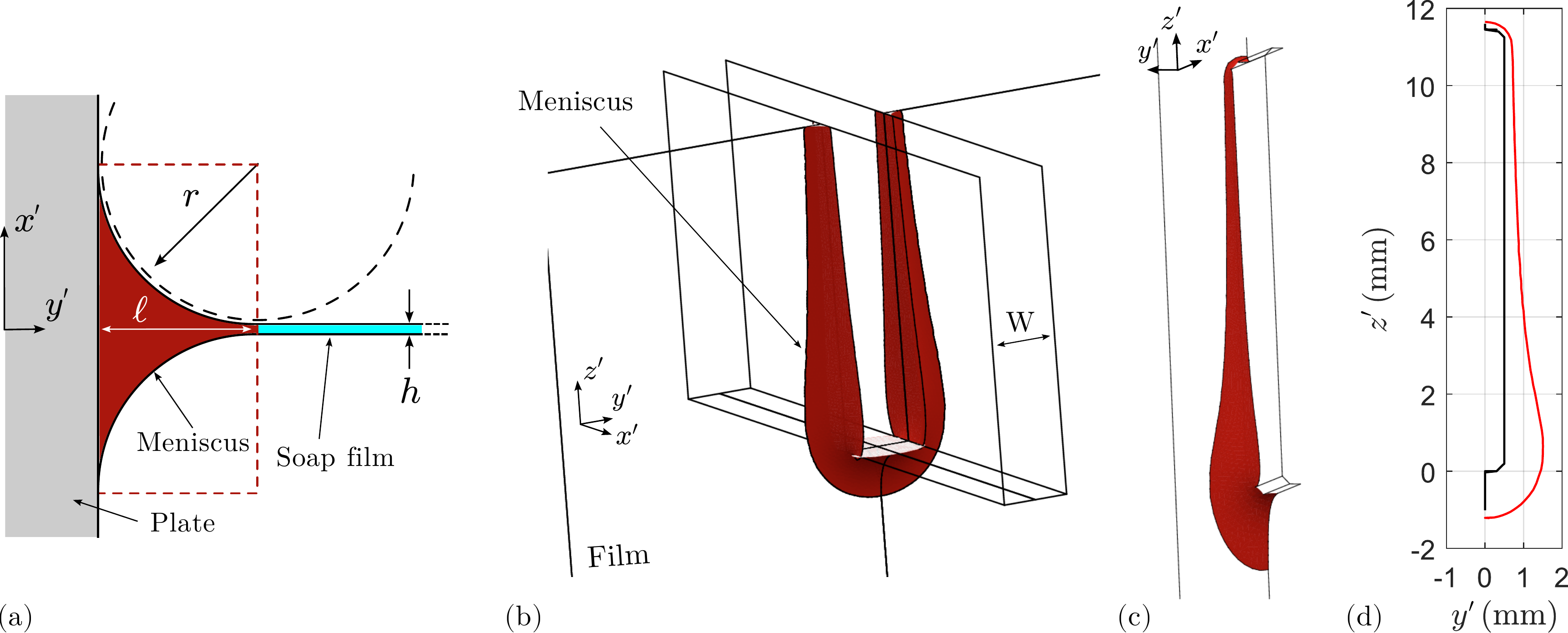}
    \caption{(a)~Cross-sectional view of the meniscus (in red) connecting the thin soap film (in cyan) to the solid plate (in gray) in the case of total wetting ($\theta=0^\circ$). The meniscus is characterized by a radius of curvature $r$ and an apparent length $\ell = r$. (b)~Example of a simulation showing a plate of width~$W$ passing through the plane of the soap film. The meniscus is shown in red in the 3D view.
    (c)~Alternative 3D view of the half-plate with a bevelled edge.
    (d)~Meniscus profile obtained by numerical simulations for a plate of height~$H=11.45$~mm, width $W=1$~mm and $\alpha=0^\circ$.
    }
    \label{Fig:simulation}
\end{figure*}

The meniscus profile along the plate is also determined using numerical simulations with Surface Evolver~\citep{brakke92}. This software is used to determine the thickness profile of the interface at equilibrium, for a given volume of liquid within the meniscus. Therefore, comparison with experiments is possible only in the hydrostatic regime (see below), where the flux from the soap film to the meniscus is negligible.

The experimental situation is reproduced numerically only for vertical plates, as illustrated in Fig.~\ref{Fig:simulation}(b). Using symmetry, only one-quarter of the domain is simulated, with mirror boundary conditions along $x=0$ and $y=0$ (see Figs.~\ref{Fig:simulation}(b), (c) and (d)). 
The density of the liquid is taken to be $\rho=10^3~{\rm kg/m}^3$, the surface tension is set to be $\gamma = 30 ~{\rm mN/m}$, and gravity is $g=9.81~{\rm m/s}^2$. Neither the thin soap films nor wetting films on the plate are present in the simulation, but their area and their influence on the shape of the meniscus are included through integrals along the meniscus-film contact lines. 
To improve numerical convergence, two small adjustments were made compared to the experimental conditions. 
First, a contact angle, $\theta$, was set along the contact lines with the soap film and with the plate. In practice we expect the value of $\theta$ to be close to $0^\circ$, but we varied it between 0 and 4$^\circ$ to probe the effect of the wetting properties. 
Second, the corner of the plate was  cut off (bevelled) at $45^\circ$ over a distance $0.2$~mm from where the corner would be (see Figs.~\ref{Fig:simulation}(c) and (d)), achieved with constraints on the position of the edge of the liquid domain in this area~\citep{brakkebook}. This reduces the turning angle at the corner (from $90^\circ$), which affects the shape of the meniscus locally. Since the interface has constant mean curvature, any change in curvature at the corner must be balanced by a change in curvature in a perpendicular direction, and hence a sharp corner gives rise to a strong dip in the contact line. The effect on the overall profile shape when different corner shapes are used is negligible (data not shown).

For each plate height $H$, the simulation starts with a small liquid volume, contact angle $\theta=8^\circ$, and no gravity; gravity is added in small steps, then the contact angle reduced in steps of $0.5^\circ$ to the required value, then the liquid volume increased, with several hundred minimization steps following each small change.  
For each value of the liquid volume, convergence to a minimum of surface energy takes only a few minutes, and we check that all eigenvalues of the Hessian of energy have the same sign~\citep{brakke96}, confirming that the liquid is in a stable configuration. The volume is then gradually increased until the meniscus shape reaches an unstable configuration, where the simulation is stopped.

\section{Steady meniscus profile}

In this section, we consider the meniscus profile when the meniscus has reached a steady state, i.e. when liquid is visibly leaking from the bottom of the plate in the form of a jet or droplets in experiments (see Fig.~\ref{Fig:setup_profile}(b)) or before numerical instability in simulations.

\subsection{Numerical benchmark\label{Sec:NumBench}}

\begin{figure}
    \centering
    \includegraphics[height=0.33\linewidth]{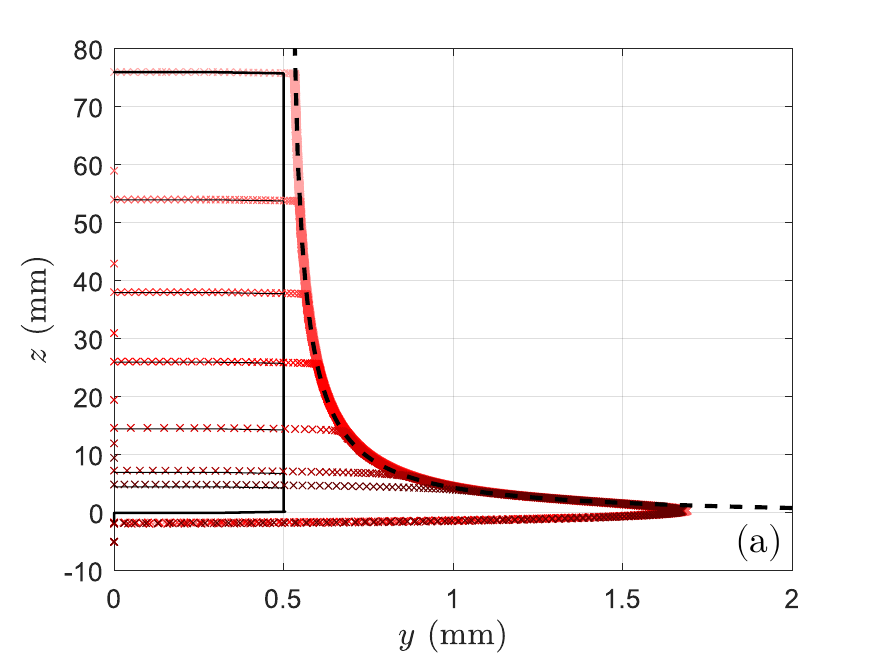}
    \includegraphics[height=0.33\linewidth]{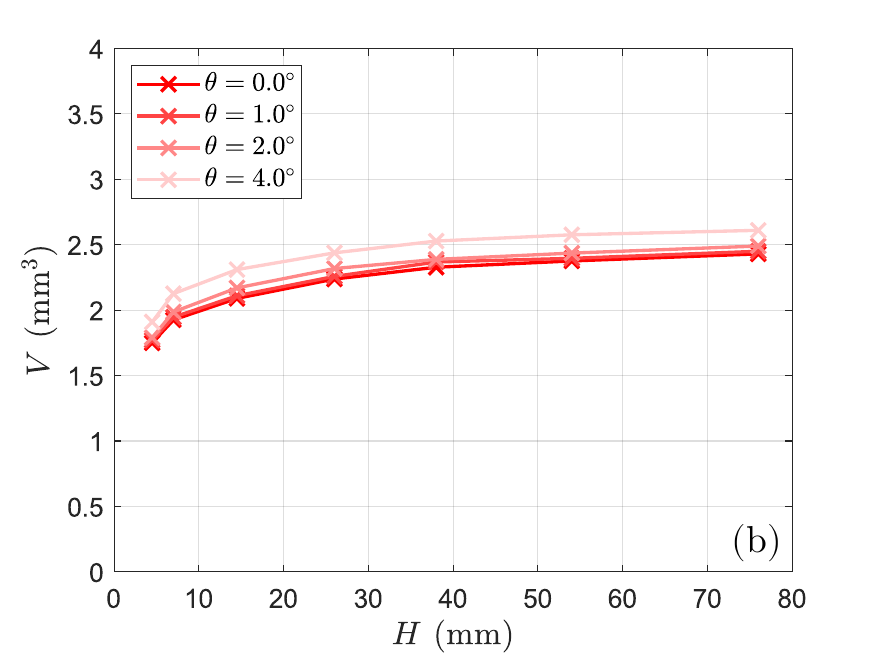}
    \caption{
Simulated menisci. 
(a)~Profiles $y(z)$ for vertical plates of various heights $H$, with fixed width $W = 1$~mm and contact angle $\theta=0^\circ$, at the maximum volume prior to the onset of instability. The profiles are shown as small red crosses, with a color gradient ranging from dark red (short plates) to light red (long plates). All profiles collapse onto a single curve; the dashed black line represents Eq.~\eqref{Eq:RmCapillary} with $y(z) = r(z) + W/2$ and $z_\infty = -1$~mm. The solid black lines indicate the positions of the plate edges. 
(b)~Maximum half-volume $V$ of the meniscus as a function of the plate height~$H$ for various contact angles~$\theta$.}
    \label{fig:V_de_H}
\end{figure}

We first validate the numerical simulations by comparing the numerical profile with the expected theoretical hydrostatic profile. At equilibrium, the shape of a vertical meniscus is determined by the balance between the Laplace pressure gradient and gravity, expressed as $\mathbf{0} = \mathbf{\nabla}(\gamma/r) + \rho \mathbf{g}$, where $\rho$ is the liquid density and $\mathbf{g}$ is the gravitational acceleration. Integrating this balance yields the variation of the radius of curvature with height $z$: \begin{equation}\label{Eq:RmCapillary} 
r(z) = \frac{\lambda_c^2}{z-z_{\infty}}, 
\end{equation} 
where $\lambda_c = \sqrt{\gamma/(\rho g)}$ is the capillary length and $z_\infty$ corresponds to the theoretical vertical position, always negative, at which the radius diverges. 

This expression allows us to compute the distance to the mid-plane, $y(z)=r(z)+W/2$, as defined in Sec.~\ref{sec:experimentalsetup}, and to compare it with the profiles obtained from the numerical simulations presented in Fig.~\ref{fig:V_de_H}(a) for a plate of width~$W = 1$~mm and various heights~$H$.  
For each value of $H$, the profile corresponds to the maximum volume reached in simulations, with larger volumes leading to unstable configurations.
Regardless of $H$, all meniscus profiles collapse onto a single curve, with deviations occurring only near the top of the plate, where $y(H)$ approaches zero. For $z > 0$, this behavior is consistent with the hydrostatic profile shown by the dashed line, corresponding to $z_\infty = -1$~mm in Eq.~\eqref{Eq:RmCapillary}. 
This result validates the numerical simulations. As expected from the definition of $r$, the two profiles no longer coincide for $z < 0$, as the curvature is no longer confined to the $x$–$y$ plane and $\ell(z)$ no longer matches $r(z)$.

We define the half-volume of the meniscus, $V$, as the volume of liquid on the right-hand side of the plate, corresponding to twice the volume calculated in the simulations from a quarter of the system. 
The maximum volume prior to the onset of instability is plotted as a function of the plate height~$H$ for various contact angles in Fig.~\ref{fig:V_de_H}(b). 
For a given contact angle, the volume increases with height but remains very weakly dependent on $H$ within our parameter range, and it saturates around $2.5$~mm$^3$ for $H \gtrsim 50$~mm. This saturation is expected, as the slope of $V(H)$ scales as $(2-\pi/2) r_H^2$, with $r_H = \lambda_c^2/(H - z_\infty)$ converging to zero as $H$ becomes very large.
For a given plate height, the volume slightly increases with the contact angle, with a maximum variation of about 10\% between $0^\circ$ and $4^\circ$.

For a given height and volume, the profiles $y(z)$ are nearly indistinguishable when varying the contact angle (data not shown). 
The effect of the angle is therefore limited to shifting the onset of instability and, consequently, the maximum volume reached before instability occurs.
For these configurations, the typical extent of the liquid in the $x$-direction below the plate and in the $y$-direction is about 1-2~mm, i.e. of the order of the capillary length, suggesting a Rayleigh–Plateau-like instability leading to drop detachment or liquid flow in experiments (a regime not captured by the simulations).

\subsection{Experimental results}

\subsubsection{Vertical plates}

\begin{figure*}
    \centering
    \includegraphics[height=0.33\linewidth]{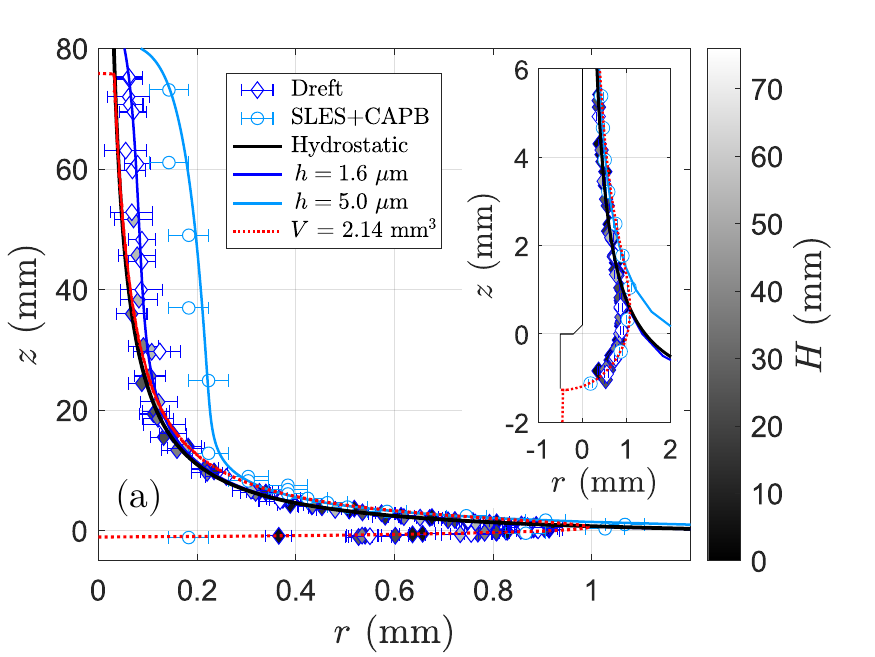}
    \includegraphics[height=0.33\linewidth]{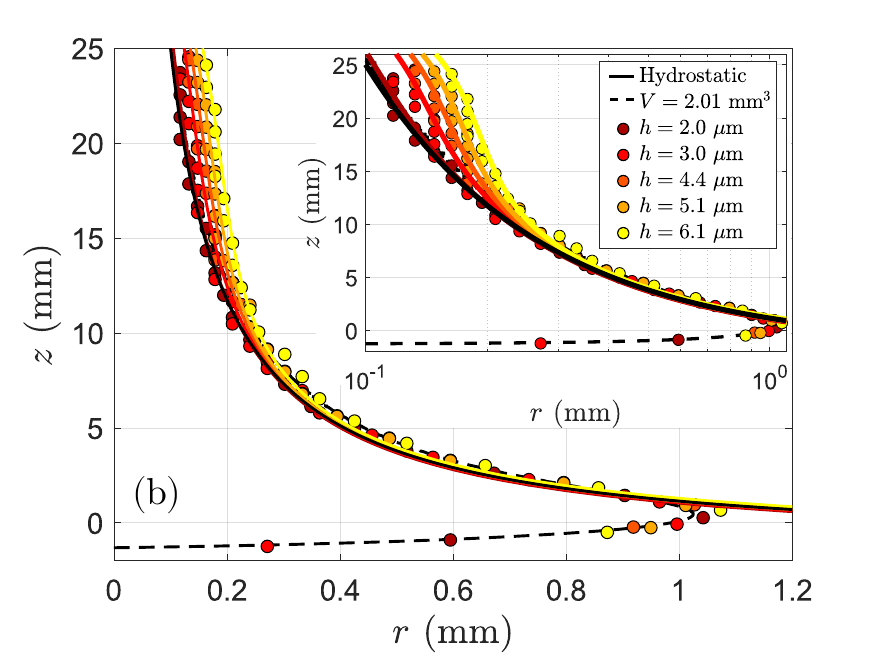}
    \caption{
(a) Profiles $r(z)$ for different plate heights $H$ ranging from 4 to 76~mm and a single film thickness $h = 1.6~\mu$m (Dreft solution, blue contours), together with a profile for the largest height and $h = 5.0~\mu$m (SLES-CAPB solution, cyan contours). The symbol color (grayscale) represents the height. The plate width is fixed at $W = 1$~mm and the inclination at $\alpha = 0^\circ$. Blue and cyan solid lines correspond to the numerical solutions of Eq.~\eqref{Eq:main_z}, imposing the radius at $z = 76$~mm and using adjusted values of the parameter $k_e$ equal to $0.3\times10^{-2}$ and $2.1\times10^{-2}$, respectively. The hydrostatic profile is shown as a black line, obtained from Eq.~\eqref{Eq:RmCapillary} with $z_\infty = -1.8$~mm. The red dotted line represents the numerical simulation for a plate of height $H = 76$~mm and width $W = 1$~mm, with a half-volume $V = 2.14$~mm$^3$.
Inset: Zoom on the bottom of the plate, with the same scale on both axes. The thin solid black line represents the plate for $z \geq 0$ and the mid-plane for $z < 0$.
(b) Profiles $r(z)$ for a single plate height $H = 26$~mm and different film thicknesses $h$ ranging from 2 to 6.1~$\mu$m, with color coding indicating $h$. Colored solid lines represent the numerical solutions of Eq.~\eqref{Eq:main_z}, imposing the radius at $z = 26$~mm and adjusting the parameter $k_e$ for each experiment (values ranging between 1.2 $\times10^{-2}$ and 3.2 $\times10^{-2}$). Again $W = 1$~mm and $\alpha = 0^\circ$. The black line shows the hydrostatic profile obtained from Eq.~\eqref{Eq:RmCapillary}, with $z_\infty = -1.5$~mm. The red dotted line represents the numerical simulation for a plate of height $H = 26$~mm and width $W = 1$~mm, with a half-volume $V = 2.01$~mm$^3$.  
}
    \label{fig:example_profile}
\end{figure*}

\begin{figure*}
    \centering
    \includegraphics[width=0.3295\linewidth]{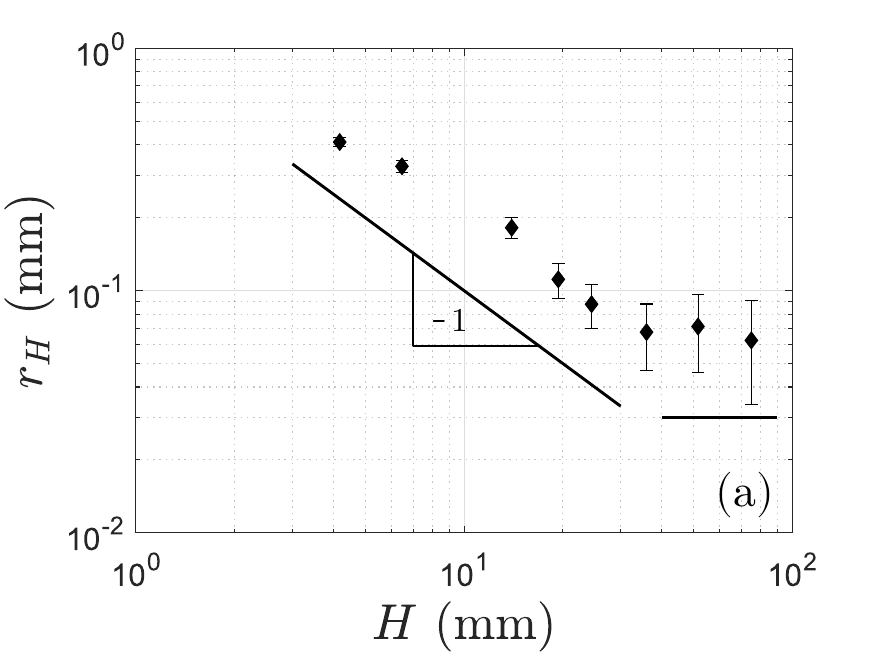}
    \includegraphics[width=0.3295\linewidth]{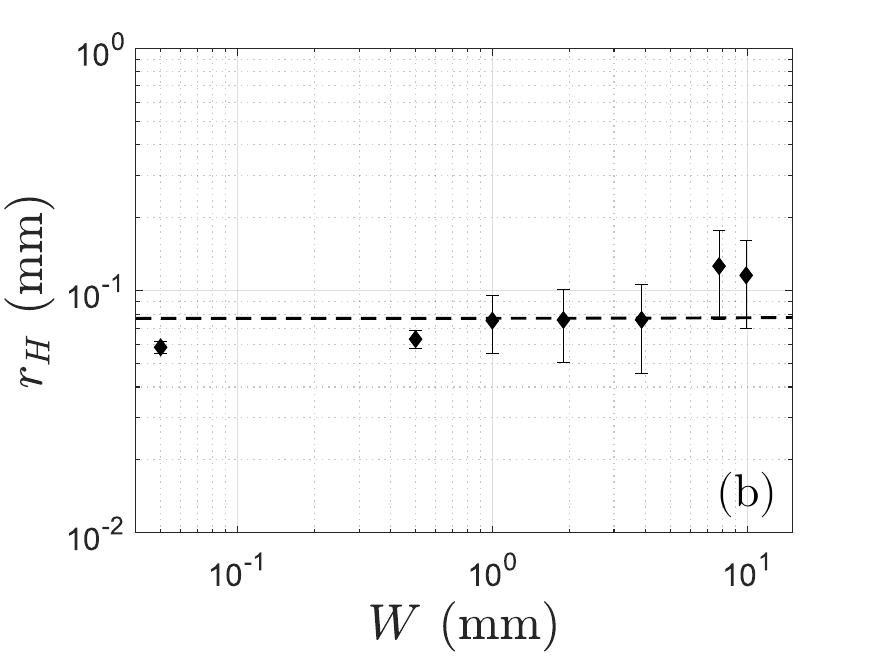}
    \includegraphics[width=0.3295\linewidth]{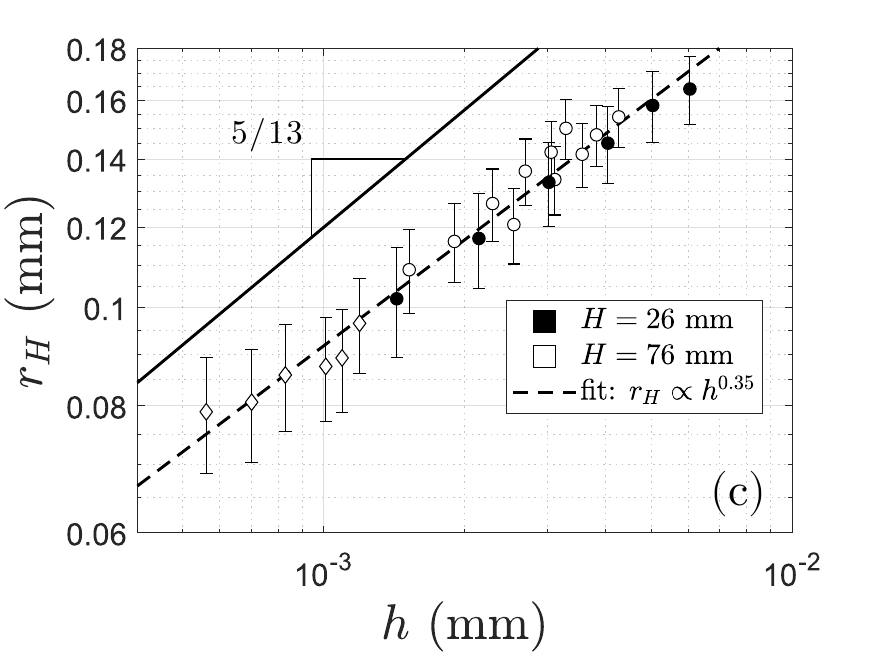}
    \caption{(a)~Meniscus curvature~$r_H$ as a function of the height of the plate~$H$, $W=1$~mm and $h=1.6~\mu$m in log-log scale.
    (b)~$r_H$ as a function of the plate width~$W$ at fixed $H=76$~mm and $h=1.17\pm0.02$~$\mu$m in log-log scale. The dashed line represents the model incorporating the linear correction in $W$.
    (c)~$r_H$ as a function of film thickness~$h$ mesured at the top of the plate in log-log scale for different type of surfactants (Dreft with diamonds and SLES + CAPB with circles) and different $H$ in color. Black dashed line shows the fit, power 0.35 (expected 5/13) on a plate with $W=1$~mm.  
    }
    \label{fig:changeparam}
\end{figure*}

We first focus on vertical plates of fixed width~$W=1$~mm, varying both their heights~$H$ and the thickness~$h$ of the surrounding soap film. Typical meniscus profiles $r(z)$ are shown in Fig.~\ref{fig:example_profile}(a). For $h=1.6~\mu$m, all profiles remarkably collapse onto the same curve (blue diamonds). When increasing the soap film thickness~$h$ to $5~\mu$m (cyan circles), we observe that the radius of curvature becomes larger at the top, while it retains the same profile at the bottom of the plate. This shows that the thickness~$h$ of the soap film is most significant at the top of the plate. This phenomenon is also clearly visible for a shorter plate in Fig.~\ref{fig:example_profile}(b), where the film thickness gradually increases from $2$ to $6.1~\mu$m.

These profiles can first be compared to the expected theoretical profile in the hydrostatic case, given by Eq.~\eqref{Eq:RmCapillary}, and the profiles obtained numerically, already presented in Fig.~\ref{fig:V_de_H}(a). Note that the half-volume~$V$ in the simulations has been adjusted to obtain the best match with the experimental profile and does not correspond to the maximum volume accessible numerically (see discussion in Section~\ref{sec:transient_results}.). Both profiles are plotted in Fig.~\ref{fig:example_profile}, respectively with a solid black line and a dotted red line. 
These profiles closely match the experimental points at the bottom of the plate, with fitted values of $z_\infty = -1.8$~mm and $z_\infty = -1.5$~mm in the theoretical profile for the long and short plates, respectively. 
Note the hydrostatic profile predicts a divergence of the radius of curvature at $z=z_\infty$, i.e. slightly below $z=0$. This divergence is not observed experimentally or numerically, since the meniscus changes its shape below the plate to form a droplet under the plate. Therefore, in the following, we only compare the profiles~$r(z)$ above $z=0$. When $z$ increases, the comparison between hydrostatic, numerical and experimental profiles remains good up to a given critical height. Above this height, the experimental points depart from the hydrostatic profile, and seems to saturate at a larger value of radius of curvature at the top of the plate. The critical height depends on the soap film thickness~$h$: the higher the thickness, the lower the critical height.

In other words, for plates with small $H$ the profile~$r(z)$ closely matches the hydrostatic profile. Therefore, at the top of the plate $r_H\approx \lambda_c^2/H$, if $z_\infty$ remains smaller than $H$. As $H$ increases, the profile~$r(z)$ at the top of the plate becomes different to the hydrostatic profile, and saturates at the top with a value of $r_H$ mainly depending on the soap film thickness~$h$. These two features are visible in Fig.~\ref{fig:changeparam}(a), where for a fixed width $W=1$~mm and thickness~$h=1.6~\mu$m, $r_H$ first decreases as $1/H$ before saturating to a constant value. Such a transition is similar to recent observation of the meniscus profile around rings placed in a similar soap film setup~\citep{vigna-brummer_flowing_2025}.

In order to probe how $r_H$ varies with $h$ and $W$, we have performed experiments where the size of the plates ($H$ and $W$) and the thickness of the film $h$ were varied. The results are shown in Figs.~\ref{fig:changeparam}(b) and (c). For a plate of height $H=76$~mm, only a very weak dependence on $W$ is observed, as shown in Fig.~\ref{fig:changeparam}(b). Furthermore, $r_H$ depends on $h$ with a power of about $0.35$, based on the fit of the data, as illustrated in Fig.~\ref{fig:changeparam}(c). These dependencies with $h$ and $W$ are further explained by the model described below, in Section~\ref{sec:model}.

\subsubsection{Inclined plates}

\begin{figure*}
    \centering
    \includegraphics[height=0.33\linewidth]{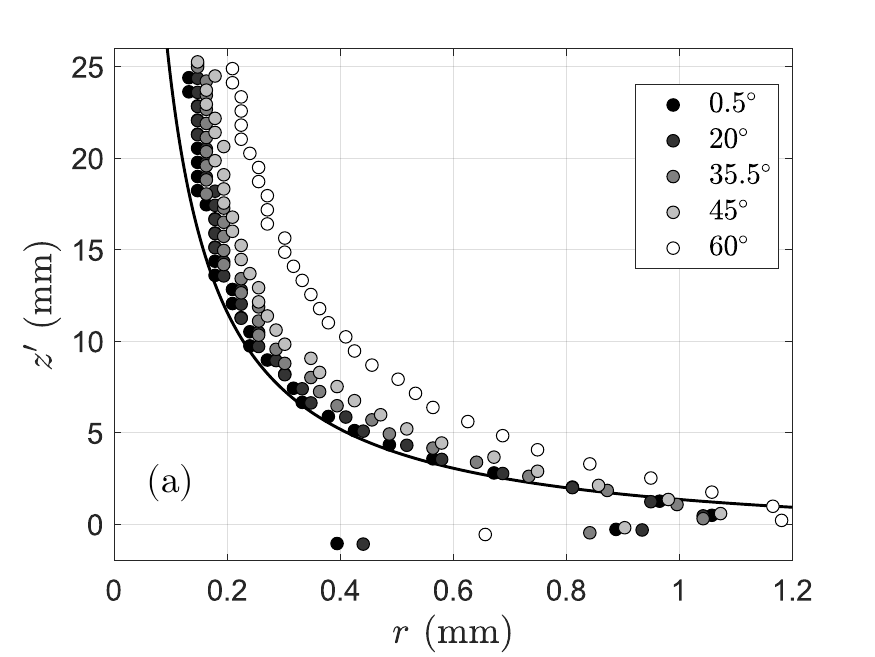}
    \includegraphics[height=0.33\linewidth]{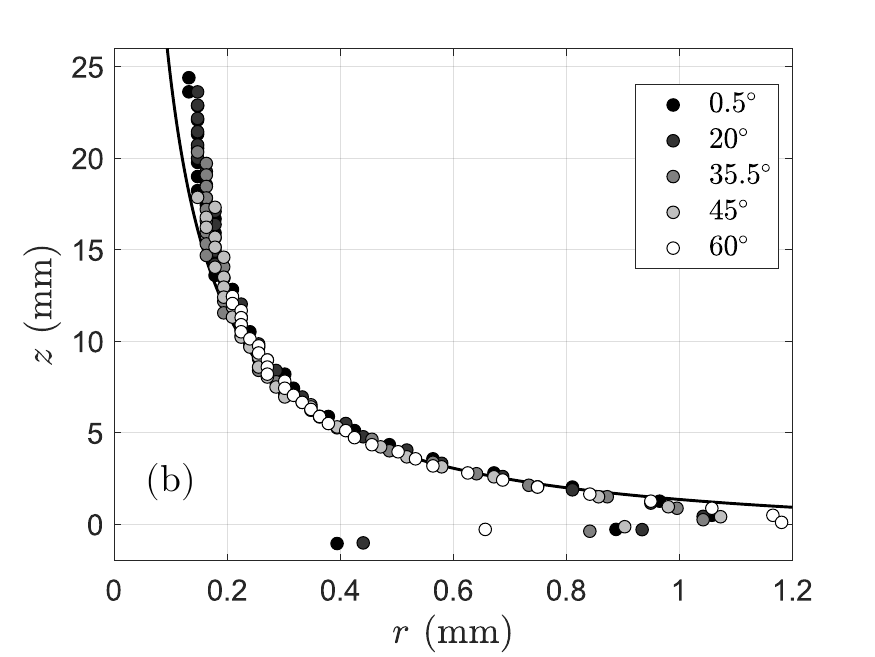}
    \caption{
(a) Profiles $r(z')$ in the plate coordinate system for different inclination angles~$\alpha$, shown using a grayscale gradient. Plate parameters are $W = 1$~mm and $H = 26$~mm. The film thickness is $h = 4.4~\mu$m. The black line represents the hydrostatic model given by Eq.~\eqref{eq:inclined}, with $z'_\infty = -1.2$~mm and $\alpha = 0^\circ$. 
(b) The same data plotted in the vertical coordinate system, $r(z)$. The black line represents the hydrostatic model given by Eq.~\eqref{eq:inclined}, with $z_\infty = -1.2$~mm.
    }
    \label{fig:sloped_profiles}
\end{figure*}

We next consider inclined plates of fixed width~$W=1$~mm and height~$H=26$~mm, varying only their inclination angle~$\alpha$. The thickness of the soap film at the top of the plate is here fixed to $h=4.4~\mu$m. Figure~\ref{fig:sloped_profiles}(a) shows the meniscus radius profiles as a function of the $z'$-coordinate along a plate 
for different inclination angles~$\alpha$. For small angles~$\alpha$, the profiles are similar to those in the vertical configuration given by Eq.~\eqref{Eq:RmCapillary}, while increasing the angle leads to a shift in the profiles, with larger radius of curvature along the plate, meaning that the meniscus becomes thicker.

For an inclined plate at an angle~$\alpha$, the theoretical profile of the hydrostatic meniscus given by Eq.~\eqref{Eq:RmCapillary} becomes
\begin{equation}
    r(z') = \frac{\lambda_c^2}{(z' - z'_\infty)\cos{\alpha}}=\frac{\lambda_c^2}{z - z_\infty},\label{eq:inclined}
\end{equation}
with $z_\infty=\cos \alpha z'_\infty$. When $\alpha$ is close to $90^\circ$, i.e. when the plate is close to horizontal, it represents a particular case where, to within an order of magnitude, $H\cos{\alpha}\sim z_\infty$. Experimentally, this makes it difficult to evaluate $z_\infty$ or leads to a considerable error in its measurement. As a consequence, we have limited our experiments to $\alpha \leq 60^\circ$. Following Eq.~\eqref{eq:inclined}, we can collapse all the profiles shown in Fig.~\ref{fig:sloped_profiles}(a) by plotting them as a function of the vertical $z$-coordinate, as illustrated in Fig.~\ref{fig:sloped_profiles}(b). Again, compared to the hydrostatic model with $z_\infty=-1.2$~mm (black line), there is good agreement near the bottom (excluding the droplet region), while for $z>15$~mm, the data deviate, as previously observed in Fig.~\ref{fig:example_profile}(a), depending on the soap film thickness~$h$. Here, $h=4.4~\mu$m, which is large enough for the deviation to be visible. Therefore, the emergence of a universal profile is still observed when $h$ is fixed, controlled by the droplet shape at the bottom, as highlighted by the invariance of the parameter $z_\infty$ in describing the lower part of the meniscus. The inclination of the plate therefore plays only a role in changing the effective height of the plate, but not the typical meniscus profile. 

\subsection{Model and comparison with data}\label{sec:model}

We now derive a model for the meniscus profile along a plate inserted into a vertical soap film. This model is similar but not identical to one derived previously~\citep{vigna-brummer_flowing_2025} for the meniscus around a ring. 
Note that we derive this model for the generic case of an inclined plate. To recover the equations and quantities derived for a vertical plate, it is therefore necessary to set $\alpha=0$. 

\subsubsection{Steady profile equation and first determination of the flux coefficient $k_e$}
 
Inside the meniscus, we consider the general case where the liquid can flow.
We work in the $(x', y', z')$ coordinate system attached to the plate, as shown in Fig.~\ref{Fig:setup_profile}(b), with $z'$ the coordinate along the plate, pointing upwards. Using the lubrication approximation, the velocity field can be expressed as $\vec{v}=-v\vec{e_{z'}}$. Assuming inertia is negligible and that the meniscus shape is unchanging and the flow steady, the Stokes equation projected along the $z'$ axis can be written as
\begin{equation}\label{Eq:Stokes}
    0 = -\frac{\partial P}{\partial z'} - \rho g \cos \alpha- \eta\left(\frac{\partial^2 v}{\partial x'^2}+\frac{\partial^2 v}{\partial y'^2}\right),
\end{equation}
where $P$ is the pressure in the meniscus. According to Laplace’s law, the pressure $P$ is given by $P = P_0 - \gamma/r$, with $P_0$ the atmospheric pressure. We can assume a no-slip boundary condition at the contact with the plate, but the velocity remains non-zero at the fluid-air interface since the soap film interface is mobile. Therefore, computing the velocity profile in the $x'-y'$ plane is very difficult. The typical average velocity in the meniscus can however be estimated from Eq.~\eqref{Eq:Stokes} as
\begin{equation}
\bar{v} \sim \frac{\gamma}{\eta} \left( \frac{\mathrm{d}r}{\mathrm{d}z'} + \frac{r^2}{\lambda_c^2}\cos \alpha\right).\label{Eq:velocity}
\end{equation}
Consequently, the average flux $Q$ in the meniscus is
\begin{equation}\label{Eq:mean_Q}
    Q = k_p\frac{\gamma}{\eta} \left( \frac{\mathrm{d}r}{\mathrm{d}z'} + \frac{r^2}{\lambda_c^2}\cos\alpha\right)r^2,
\end{equation}
where $k_p$ is a constant that accounts for the geometry and the permeability of the meniscus. For a meniscus with the cross-sectional shape considered here, this constant has been estimated by numerical simulations for different interface mobilities~\citep{drenckhan_rivulet_2007}. For large mobility, which is the case here, the constant reaches a plateau at the value~$k_p=2.73 \times 10^{-2}$. 

At steady state, conservation of mass between the liquid pumped from the film to the meniscus, as in Eq.~\eqref{Eq:linear_flux}, and the flow within the meniscus driven by gravity, leads to the balance equation
\begin{equation}\label{Eq:bilan}
    \frac{\mathrm{d}Q }{\mathrm{d}z'} = -q.
\end{equation} 
Substituting for $Q$ and $q$ from  
Eqs.~\eqref{Eq:mean_Q} and \eqref{Eq:linear_flux}, we get
\begin{equation}\label{Eq:main}
    r^{3/2}\frac{\mathrm{d}}{\mathrm{d}z'}\left( r^4 \cos \alpha+ \lambda_c^2r^2\frac{\mathrm{d}r}{\mathrm{d}z'}\right) = -\frac{k_e}{k_p}\lambda_c^2h^{5/2},
\end{equation}
or using the $z$-coordinate:
\begin{equation}\label{Eq:main_z}
    r^{3/2}\frac{\mathrm{d}}{\mathrm{d}z}\left( r^4+ \lambda_c^2r^2\frac{\mathrm{d}r}{\mathrm{d}z}\right) = -\frac{k_e}{k_p}\frac{\lambda_c^2 h^{5/2}}{\cos^2\alpha}.
\end{equation}
This second order differential equation requires two boundary conditions. One corresponds to  zero flux at the top of the plate, which imposes 
\begin{equation}
\frac{\text{d}r}{\text{d}z} (z=H\cos\alpha) + \frac{r^2(z=H\cos\alpha)}{\lambda_c^2}=0 . \label{eq:BC_flux}
\end{equation}
For the second boundary condition, in order to compare with experiments, we impose 
\begin{equation}
r(z = H\cos\alpha) = r_H
\end{equation}
to match the uppermost data points. This choice fixes $\displaystyle\frac{\text{d}r}{\text{d}z}\left(z=H\cos\alpha\right)$ and allows the numerical resolution of the full profile for fixed parameters in Eq.~\eqref{Eq:main_z}. The only remaining free parameter is $k_e$, which is adjusted to fit the experimental data for $z > 0$. As shown in Fig.~\ref{fig:example_profile}, this procedure yields very good agreement between the experiments and the model, allowing a reliable determination of $k_e$ for each experiment.
These values are reported in Fig.~\ref{fig:rH_de_rg_normby_rs}(a), where $k_e$ is found to be relatively constant across all experiments. Averaging over all measurements yields a first estimate of $k_e = (2.3 \pm 0.4)\times10^{-2}$.

Note that the lower part of the meniscus does not follow the trend predicted by the model, consistent with the fact that $r$ defined in experiments no longer represents the radius of curvature for $z < 0$. Nevertheless, we continue to denote by $z_\infty$ the position at which the radius of curvature is expected to diverge. Regardless of the regime considered, $z_\infty$ is found to lie between approximately -1 and -2~mm, remaining small compared with the heights of the plates used throughout this study.

\subsubsection{Prediction of the meniscus curvature $r_H$ and second determination of the flux coefficient $k_e$}

\begin{figure*}
    \centering
    \includegraphics[height=0.27\linewidth]{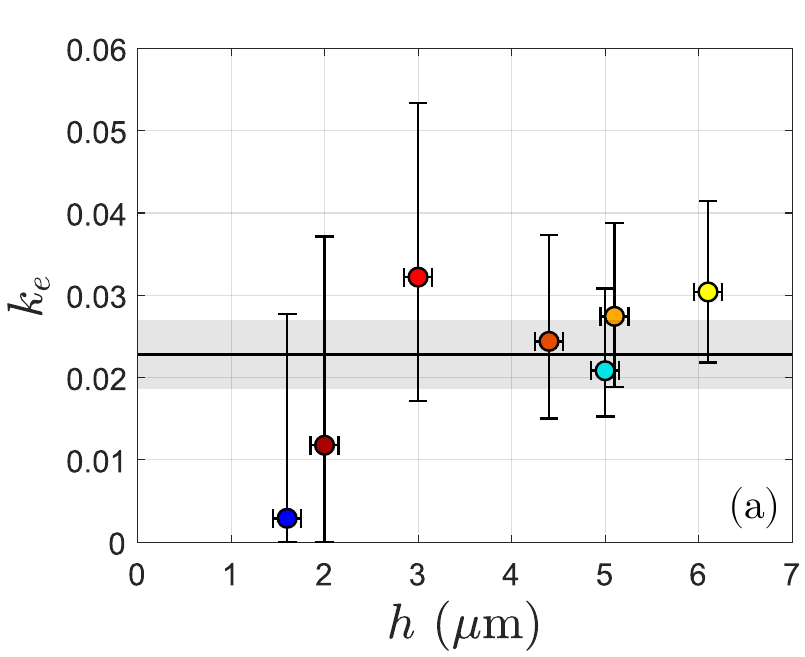}
    \includegraphics[height=0.27\linewidth]{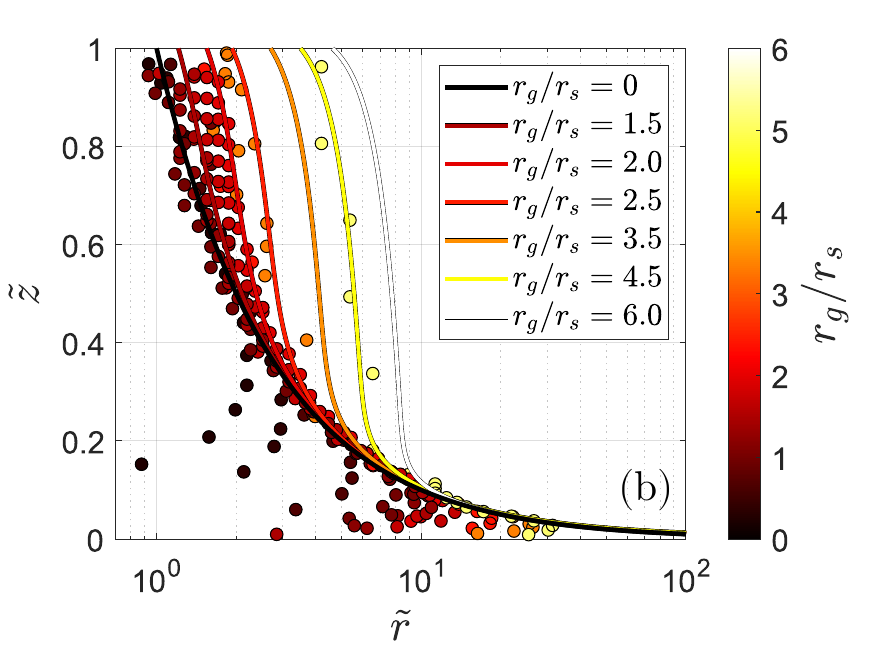}
    \includegraphics[height=0.27\linewidth]{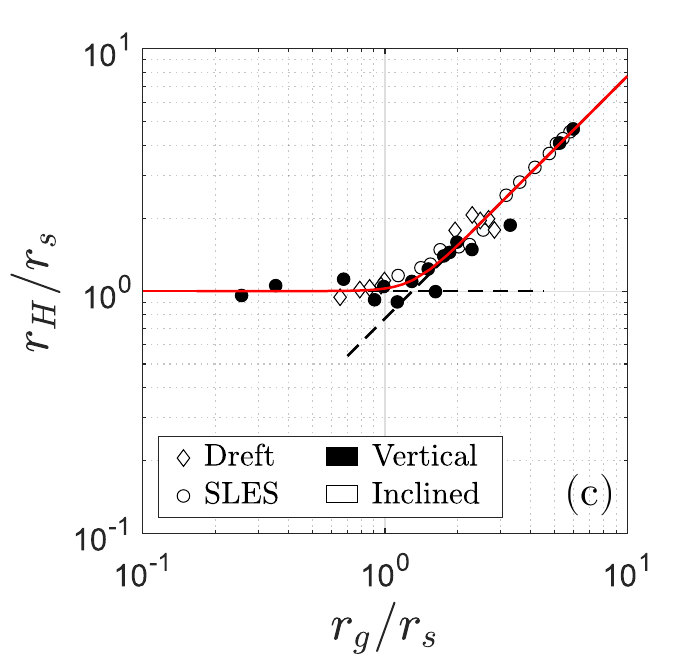}
    \caption{
(a)~Flux coefficient $k_e$ obtained from the best fits of the model (Eq.~\eqref{Eq:main_z}) to the experimental data shown in Figs.~\ref{fig:example_profile}(a,b), using the same color coding. Error bars indicate the associated uncertainties. 
A weighted average over all values yields $k_e = (2.3 \pm 0.4)\times10^{-2}$, where the weights are taken as the inverse square of the error bars. 
(b)~Normalized meniscus profiles $\tilde{r}(\tilde{z})$ for various ratios of $r_g/r_s$, as indicated by the color bar. Colored circles denote experimental data, while solid colored lines represent the solutions of Eq.~\eqref{Eq:main_z_dimensionless}. The limit $r_g/r_s = 0$ (black line) corresponds to the hydrostatic case described by Eq.~\eqref{Eq:RmCapillary}, which, in normalized form, is $\tilde{r} = 1/\tilde{z}$. 
Note that data points lying below the hydrostatic limit correspond to positions where $z_\infty<z < 0$, and $r$ no longer represents the meniscus radius of curvature.
(c)~Meniscus curvature $r_H/r_s$ as a function of $r_g/r_s$ in a log–log representation. The red curve corresponds to the model prediction. The horizontal black dashed line represents the hydrostatic limit $r_H/r_s = 1$, while the sloped dashed line indicates the asymptotic regime $r_H/r_s = 0.77\,r_g/r_s$. In this approach, a single value of $k_e$ is tuned to determine $r_g$ in the experiments such that the data follow the relation $r_H/r_s = 0.77\,r_g/r_s$ in the large $r_g/r_s$ limit. The best fit yields $k_e = (2.7 \pm 0.5)\times10^{-2}$.}
    \label{fig:rH_de_rg_normby_rs}
\end{figure*}

A full prediction can be obtained by replacing the second boundary condition with $r(z_\infty) \rightarrow \infty$, rather than relying on the uppermost experimental data points. Equation~\eqref{Eq:main_z}, together with its two boundary conditions, can be made dimensionless by using the meniscus radius of curvature in the hydrostatic regime at the top of the plate, $r_s=\lambda_c^2/(H \cos \alpha -z_\infty)$ (see Eq.~\eqref{Eq:RmCapillary}). We thus define the dimensionless radius $\tilde{r} = r / r_s$ and we normalize the $z$-coordinate through $\tilde{z}=(z -z_\infty)/(H \cos \alpha -z_\infty)$. 
This leads to the dimensionless equation
\begin{equation}\label{Eq:main_z_dimensionless}
    \tilde{r}^{3/2}\frac{\mathrm{d}}{\mathrm{d}\tilde{z}}\left( \tilde{r}^4+ \tilde{r}^2\frac{\mathrm{d}\tilde{r}}{\mathrm{d}\tilde{z}}\right) = -\frac{k_e}{k_p}\frac{\lambda_c^4 h^{5/2}}{r_s^{13/2} \cos^2\alpha} \equiv -\left(\frac{r_g}{r_s}\right)^{13/2},
\end{equation}
where the quantity~$r_g$ is defined as
\begin{equation}
r_g=\left(\frac{k_e}{k_p}\right)^{2/13}\left(\frac{\lambda_c}{\sqrt{\cos{\alpha}}}\right)^{8/13}h^{5/13}.\label{eq:rg}
\end{equation}
The first boundary condition of this second order differential is still given by no flow at the top of the plate, i.e. $Q(z=H\cos\alpha)=0$ (see Eq.~\eqref{eq:BC_flux}), written more explicitly as
\begin{equation}
           \frac{\textrm{d}\tilde{r}}{\textrm{d}\tilde{z}}(\tilde{z}=1) + \tilde{r}^2(\tilde{z}=1)=0.\label{eq:BC}
\end{equation} 
The second and new boundary condition is given by the divergence in $z=z_\infty$, i.e 
\begin{equation}
\tilde{r}(\tilde{z}=0)\rightarrow \infty .
\end{equation}
Equation~\eqref{Eq:main_z_dimensionless} is solved numerically with its two boundary conditions, using $r_g/r_s$ as a control parameter (Fig.~\ref{fig:rH_de_rg_normby_rs}(b)).
For $r_g/r_s = 0$, the hydrostatic profile $\tilde{r} = 1/\tilde{z}$ is recovered, which is the dimensionless counterpart of Eq.~\eqref{eq:inclined}. Only small deviations from this profile are observed for $r_g/r_s < 1$. As this parameter increases beyond unity, the upper part of the profile is progressively modified while the lower part remains essentially unchanged; moreover, the larger $r_g/r_s$, the lower along the plate the transition occurs. 
In this dimensionless representation, the experimental data are in quantitative agreement with the predicted profiles for the corresponding values of $r_g/r_s$, except for the data points lying below the hydrostatic limit. These points correspond to positions where $z_\infty < z < 0$, for which $r$ no longer represents the meniscus radius of curvature.

The condition $r_g/r_s \sim 1$ thus defines the onset of a “flowing” meniscus, in which the exchange flow significantly alters the quasistatic meniscus shape. From the definition of $r_s$, this criterion can be expressed in terms of a critical plate height $H_c$, above which the flowing regime becomes dominant, or equivalently the position where the transition occurs for a long plate with $H>H_c$:
\begin{equation}\label{Eq:H_c}
H_c \cos \alpha = z_\infty + \frac{\lambda_c^2}{r_g}.
\end{equation} 
With $-z_\infty$ of order 1–2~mm, $H_c$ can be approximated as $\lambda_c^2/(r_g \cos \alpha)$ within our parameter range. For $h = 1.6~\mu$m and $\alpha = 0$, this yields $H_c \approx 25$~mm, which is fully consistent with the value shown in Fig.~\ref{fig:example_profile}(a) for the blue diamonds. As $h$ increases, $r_g$ increases and $H_c$ decreases, consistent with the cyan circles in the same figure. This trend is also visible in Fig.~\ref{fig:example_profile}(b), where increasing film thickness $h$ leads to smaller values of $H_c$ and correspondingly weaker deviations from the hydrostatic profile.

From the predicted profiles, we can also extract the radius of curvature at the top of the plate, $r_H$. As shown in Fig.~\ref{fig:rH_de_rg_normby_rs}(c), $r_H = r_s$ in the limit $r_g/r_s \ll 1$, as expected from the hydrostatic solution, while in the opposite limit $r_g/r_s \gg 1$ we find $r_H = 0.77\,r_g$. 

We now compare the experimental measurements of $r_H$ with the predictions of the model, for both vertical and inclined plates. For each experiment, this comparison requires calculating $r_s = \lambda_c^2/(H \cos \alpha - z_\infty)$ and the characteristic radius $r_g$ given by Eq.~\eqref{eq:rg}. The quantity $r_s$ is evaluated using the imposed experimental parameters and the value of $z_\infty$ obtained by fitting the lower part of the profile, between $z = 0$ and $H_c$, with Eq.~\eqref{eq:inclined}. Note that the precise determination of $z_\infty$ has little influence on the results, since $-z_\infty$ remains smaller than $H \cos \alpha$ in all cases. Determining $r_g$ requires specifying a value for $k_e$, which is treated here as a global fitting parameter common to all experiments. 
The comparison is performed by plotting $r_H / r_s$ as a function of $r_g / r_s$ (Fig.~\ref{fig:rH_de_rg_normby_rs}(c)). Regardless of the chosen value of $k_e$ for the experimental data, $r_H / r_s \propto r_g / r_s$ in the large-$r_g / r_s$ limit. We therefore tune $k_e$ so that the proportionality coefficient matches the model prediction, which yields $k_e = (2.7 \pm 0.5)\times10^{-2}$. This provides a second determination of $k_e$, seemingly insensitive to the plate geometry (inclination, height, and width). Once $k_e$ is fixed, we observe very good agreement between experiments and the model for both vertical and inclined plates, indicating that $k_e$ is independent of $\alpha$ within the precision of our measurements.

This framework allows us to understand the trends observed in Fig.~\ref{fig:changeparam}. The crossover around $H \simeq 25-30$~mm in Fig.~\ref{fig:changeparam}(a) corresponds to the condition $r_g = r_s$. The power-law decrease of $r_H$ with an exponent of $-1$ at small $H$, together with the saturation of $r_H$ at large $H$, correspond to the quasistatic and flowing regimes, respectively.
The data for $r_H$ as a function of the plate width $W$ in Fig.~\ref{fig:changeparam}(b) lie in the flowing regime, with $r_g > r_s$. The observed independence of $r_H$ with respect to $W$ can be attributed to the fact that $W$ does not enter the expression for $r_g$. Note, however, that very long plates could introduce an additional source term associated with liquid collected by capillary suction on the upper horizontal part of the plate, thereby modifying the zero-flux condition at $z = H$ assumed in the model. Such an effect does not appear to be significant over the range of $W$ explored here. 
Finally, the power law observed between $r_H$ and $h$ in Fig.~\ref{fig:changeparam}(c), with a fitted exponent of 0.35, is consistent within 5\% with the exponent $5/13 \approx 0.385$ (Eq.~\eqref{eq:rg}) expected in the regime $r_g \gtrsim r_s$ when $r_H \propto r_g$.

\section{Transient dynamics: growing meniscus}

\begin{figure*}
    \centering
    \includegraphics[width=0.75\linewidth]{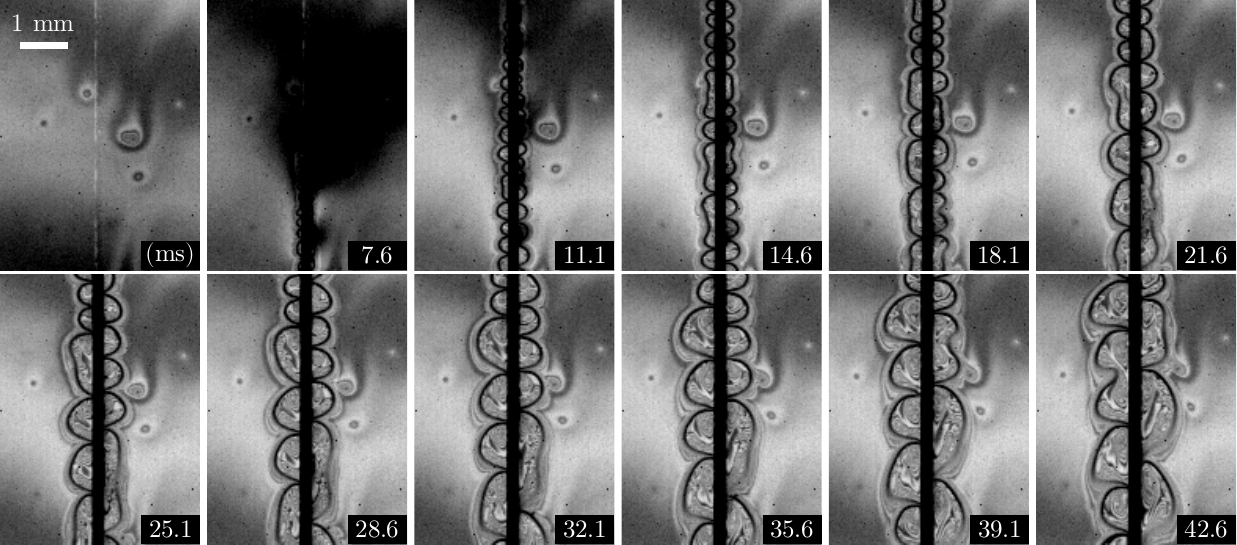}
      \caption{High-speed black-and-white images showing the initial stages of contact between a vertical glass plate ($W=150 \ \mu$m) and a soap film. The time interval between successive frames is 3.5~ms (timestamps in the bottom-right corners are given in ms). The corresponding movie is available in the Supplementary Material (movie 3).
    }
    \label{Fig:first_contact}
\end{figure*} 

All previous measurements were taken in the steady state, wherein the meniscus profile is independent of time. In this section, we focus on the transient regime, i.e. on the first moments after the plate is introduced in the film and before the meniscus reaches its stationary shape. 

Figure~\ref{Fig:first_contact} shows the very first instants just after the plate has been introduced in the film, captured using a high-speed camera. The full video is available in the Supplementary Material. The first frame corresponds to the system just before contact. At $t=7.6$~ms, contact is established, causing a local deformation of the soap film; the resulting dark regions indicate areas where the film curvature prevents reflections from being captured by the camera. At the bottom-left of the plate, the onset of TFE nucleation is already visible. From $t=11.1$~ms, TFEs are seen to grow all along the plate and continue to expand throughout the remainder of the sequence. Several coalescence events are visible, notably between $t=35.6$ and 39.1~ms near the top-right edge of the plate. Finally, an upward vertical shift is observed: once the TFEs reach a sufficient size, buoyancy effects overcome other forces, causing them to migrate upward against gravity. Note that, during this short sequence, the meniscus formed between the plate and the film is very small ($r \approx 45~\mu$m) and does not grow significantly.

\begin{figure*}
    \centering
    \includegraphics[width=0.85\linewidth]{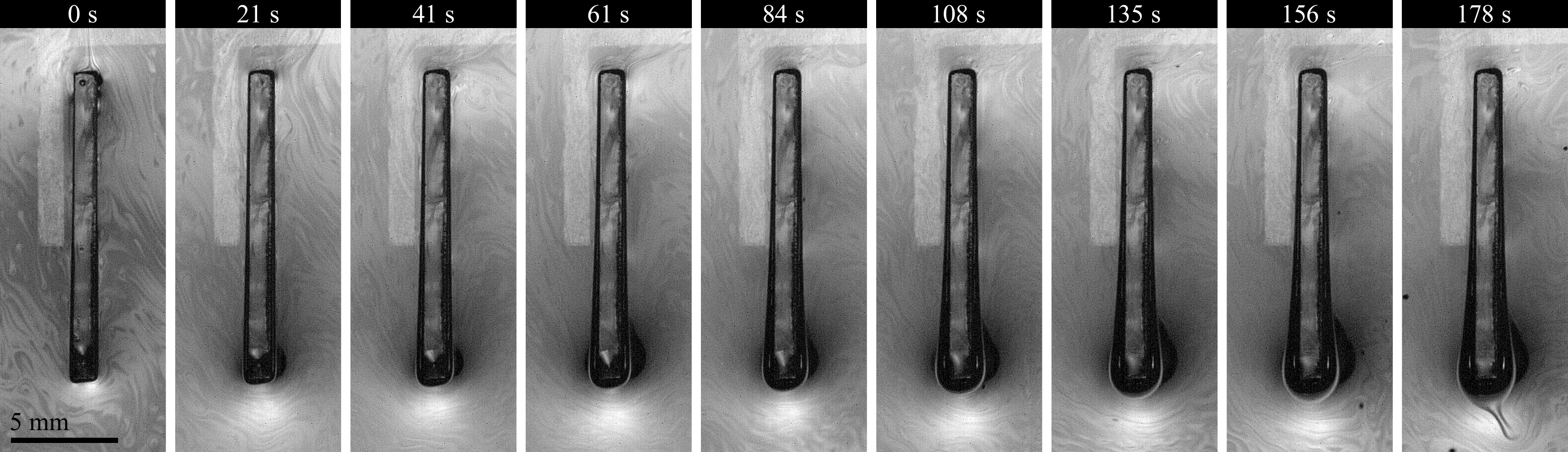}
    \caption{Image sequence showing the long-time transient dynamics of a vertical plate ($H = 14.5$~mm, $W = 1$~mm) in a vertical soap film of thickness $h = 1.6~\mu$m. The corresponding movie is available in the Supplementary Material (movie 4).}
    \label{Fig:meniscus_growth}
\end{figure*}

Since the initial dynamics of the first TFEs produced after the contact is very fast—typically occurring in less than 50 ms— we here mainly focus on longer time scales where the meniscus grows, progressively filled by liquid from the soap film, over typical durations ranging from 50 to 1500 s (typical example in Fig.~\ref{Fig:meniscus_growth}). During that stage, TFEs are continuously produced and advected towards the top of the plate. However, in contrast to the steady-state case, the meniscus does not leak at the bottom. 

\subsection{Experimental results\label{sec:transient_results}}

\begin{figure*}
    \centering
    \includegraphics[height=0.37\linewidth]{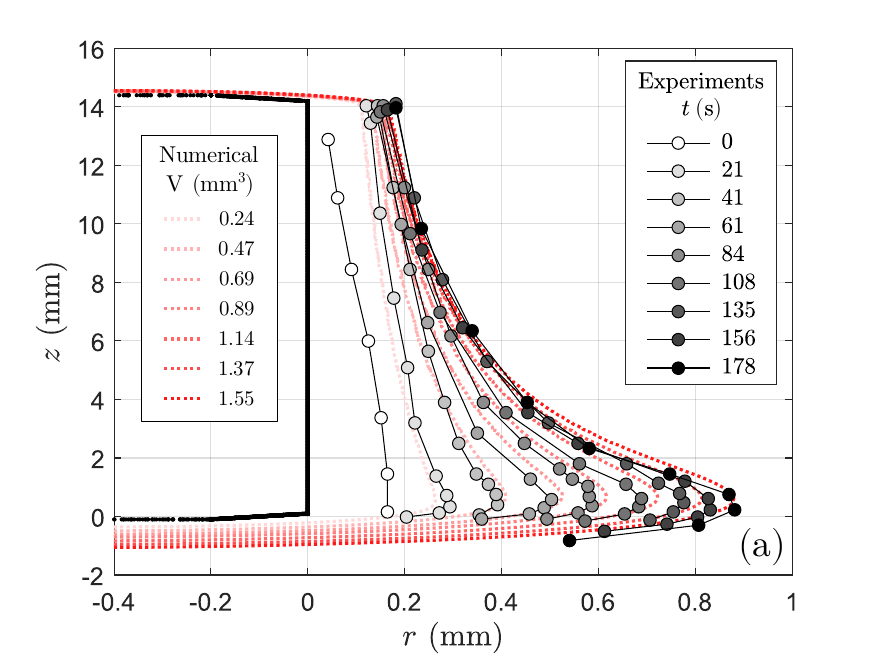}
    \includegraphics[height=0.37\linewidth]{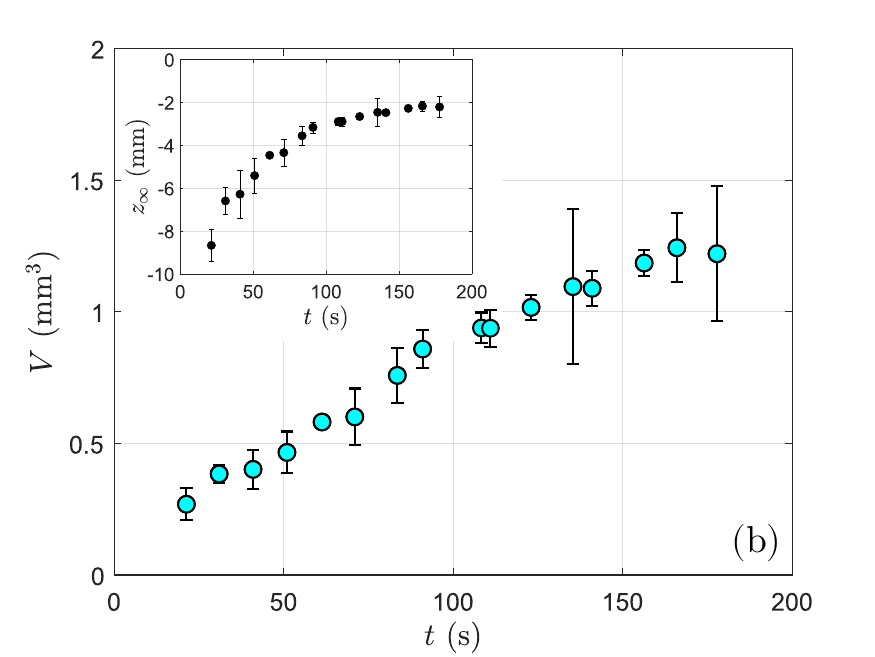}
    \caption{(a)~Meniscus profile~$r(z)$ for a plate of height~$H=14.5$~mm and width~$W=1$~mm, in a soap film of thickness~$h=1.6~\mu$m. Black dots indicate the positions of the plate, with the bottom-right edge centered at the coordinate~$(0,0)$. Circles in grayscale represent experimental profiles at successive times after immersion in the soap film, from snapshots in Fig.~\ref{Fig:meniscus_growth}, where $t=0$ corresponds to the first instant of contact ($\pm \ 0.3$ s). Dotted lines in shades of red correspond to numerical simulations for different half-volumes~$V$. The corresponding movie is available in the Supplementary Material (movie 5).
    (b) Time evolution of the liquid half-volume~$V$ from the experiments shown in panel (a). Inset: the corresponding value of $z_\infty$ versus time.}
    \label{fig:filling_profile}
\end{figure*}

We first focus on a vertical plate of relatively small height, $H = 14.5$~mm, with width $W = 1$~mm, corresponding to the one shown in Fig.~\ref{Fig:meniscus_growth}. The plate is inserted into a soap film of thickness $h = 1.6~\mu$m. Figure~\ref{fig:filling_profile}(a) shows the meniscus profiles at different times after the plate is introduced into the soap film at $t = 0$. 
The transient typically lasts about 180~s, until the meniscus begins to leak and its profile stabilizes. For these conditions, $H_c = 22$~mm, and we therefore expect the steady meniscus profile to be fully described by the hydrostatic solution given by Eq.~\eqref{Eq:RmCapillary}.  
During this transient, the radius~$r$ increases at a given height~$z$, indicating that the meniscus volume increases with time. 
These experimental profiles are compared in the same figure with hydrostatic profiles obtained from numerical simulations using identical plate parameters and varying meniscus volumes. Except at the initial time, each experimental profile at a given time closely matches a numerical profile corresponding to a specific volume.  
This comparison allows us to estimate the experimental meniscus half-volume $V(t)$ and the position $z_\infty(t)$, as shown in Fig.~\ref{fig:filling_profile}(b). The detailed procedure consists of fitting each experimental profile (in the region where $z > 0$) with Eq.~\eqref{Eq:RmCapillary} to extract $z_\infty(t)$. Using numerical simulations, we then determine the corresponding volume for each value of $z_\infty$ in order to reconstruct the experimental evolution of $V(t)$. 
Several remarks can be made. 
First, the comparison suggests that, during most of the transient, the experimental profiles also correspond to the quasistatic regime and are well described by Eq.~\eqref{Eq:RmCapillary}. 
The entire profile can therefore be characterized by a single parameter, $z_\infty(t)$: as the half-volume increases, the meniscus thickens at a given $z$, or equivalently $|z_\infty|$ decreases.  
Second, the profile at $t = 0$ cannot be described by a hydrostatic profile, indicating that, at early times, the liquid within the meniscus is not in quasistatic equilibrium. 
Third, a discrepancy is observed in the steady-state values reached at the end of the transient. While the experiments yield $V = 1.22 \pm 0.26$~mm$^3$ and $z_\infty = -2.2 \pm 0.5$~mm in the steady state, the numerical simulation with $\theta=0^\circ$ reaches a larger half-volume, $V = 2.09$~mm$^3$, with $z_\infty = -1.2$~mm. This difference is likely due to experimental perturbations—such as air currents, film motion, and/or marginal regeneration effects—which can destabilize the meniscus at its lower end and trigger leakage at earlier times.

\begin{figure*}
    \centering
    \includegraphics[height=0.33\linewidth]{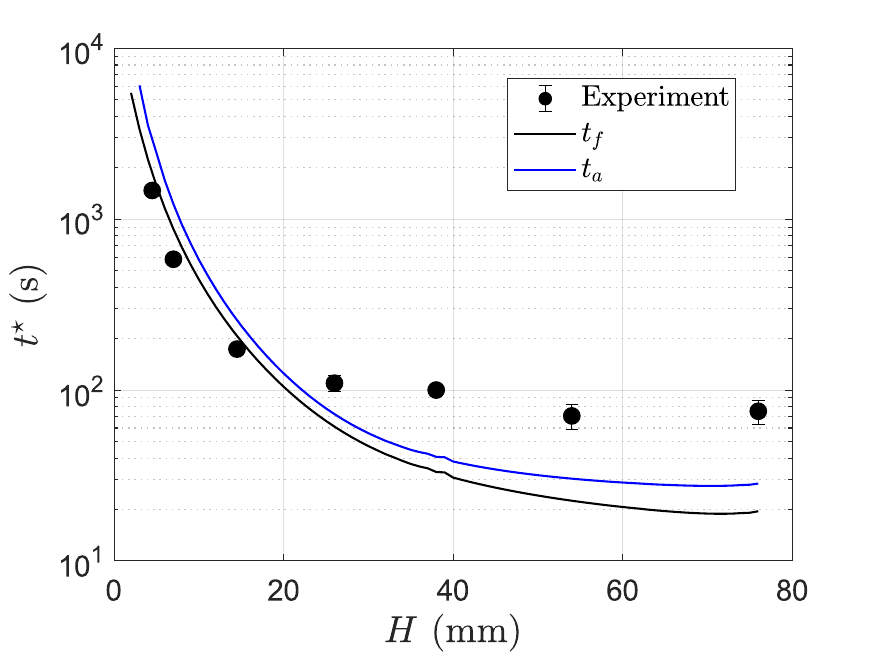}
    \caption{Transition time~$t^\star$ as a function of the height of the plate~$H$ plotted on a semi-log scale. Experiments are shown with black dots and error bars. The black line represents the filling time~$t_f$ (Eq.~\eqref{Eq:filling_time_small}) while the blue line shows the advection time~$t_a$ (Eq.~\eqref{Eq:advection}).}
    \label{fig:filling_time}
\end{figure*}

We also examined experimentally the influence of the plate height~$H$ on the transition time~$t^\star$, which marks the transition between a growing and a steady meniscus profile. It is inferred from the time at which liquid begins to leak from the bottom of the meniscus.
Figure~\ref{fig:filling_time} shows the time~$t^\star$ for different plate heights~$H$, represented by the black dots. For short plates, the transition time is large (for instance, a plate of height $H=4.5$~mm requires nearly $25$~min). As the plate height increases, the transition time decreases and appears to approach a constant value, slightly above $1$~min. This seems to indicate that, beyond a given height, the transition time no longer depends on the plate height~$H$ within the range explored in this study. Note that, in the case of large~$H$, {in particular when $H>H_c$}, the numerical simulations cannot be used to compare with the experimental profiles since the profiles are not expected to follow the hydrostatic prediction.

\subsection{Model}

\subsubsection{General case}

We now rationalize the different features observed in Figs.~\ref{fig:filling_profile} and~\ref{fig:filling_time} by considering a model for the growth of the meniscus. For $t < t^\star$, the meniscus radius of curvature depends on both space and time, $r(z,t)$. The temporal variation of an elementary meniscus volume, of cross-sectional area~$A(z,t)$ and infinitesimal height $\mathrm{d}z$, results from two contributions: the liquid exchange with the soap film, $q(z,t) \mathrm{d}z$, which feeds the meniscus, and the difference in flow rate between the top and bottom of this elementary volume, given by $(\partial Q/\partial z) \mathrm{d}z$. This balance leads to the following evolution equation:
\begin{equation}\label{Eq:bilan_temporal}
    \frac{\partial A}{\partial t} = q(z,t)+\frac{\partial Q}{\partial z}.
\end{equation}

This equation is similar to Eq.~\eqref{Eq:bilan}, but now includes a temporal term. Within the meniscus, if fluid drainage is weaker than the injection from the film, the meniscus grows; conversely, if drainage exceeds injection, it shrinks. Although Eq.~\eqref{Eq:bilan_temporal} is strictly valid only for $z > 0$, a global volume balance can nevertheless be written at the scale of the whole meniscus, including the part below $z=0$, provided that there is no source term at the top, no liquid loss at the bottom during the transient, and that the only influx arises from liquid exchange along the plate. Under these assumptions, the evolution of the meniscus half-volume obeys
\begin{equation}\label{Eq:exchange_flux}
    \frac{\mathrm{d}V}{\mathrm{d}t} = \int_{0}^Hq(z,t)\ \mathrm{d}z.
\end{equation}

\subsubsection{Short plates and third determination of the flux coefficient $k_e$}

\begin{figure*}
    \centering
    \includegraphics[height=0.33\linewidth]{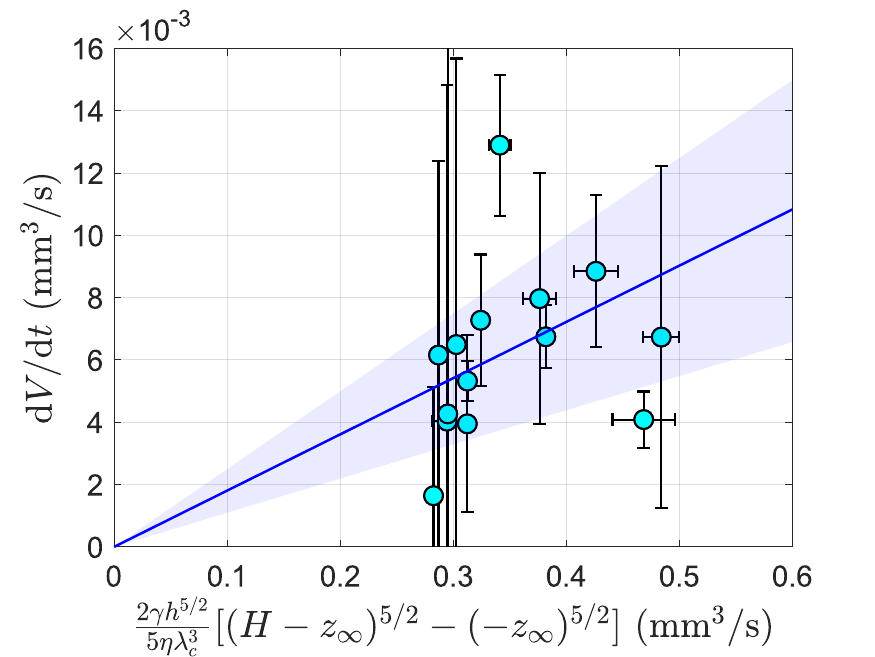}
    \caption{Measured flow rate $\mathrm{d}V/\mathrm{d}t$ as a function of the scaling prediction from Eq.~\eqref{Eq:dVdt}. The blue line shows the weighted fit resulting in $k_e=(1.8\pm0.7)\times10^{-2}$, with the shaded area representing the error bounds.}
    \label{fig:Volume_and_Flux}
\end{figure*}

We now consider the case of a small plate height, as illustrated in Fig.~\ref{fig:filling_profile}. Since the transient meniscus profiles~$r(z,t)$ in experiments coincide with the hydrostatic profiles obtained in simulations, we can use the expression for $r(z,t)$ given by Eq.~\eqref{Eq:RmCapillary}, with the time dependence entering solely through the parameter $z_\infty(t)$.
Combining Eqs.~\eqref{Eq:exchange_flux}, \eqref{Eq:linear_flux} and \eqref{Eq:RmCapillary} leads to
\begin{equation}\label{Eq:dVdt}
    \frac{\mathrm{d}V(t)}{\mathrm{d}t} = k_e\frac{2}{5}\frac{\gamma}{\eta}\frac{h^{5/2}}{\lambda_c^3}\left[(H-z_{\infty}(t))^{5/2} - (-z_{\infty}(t))^{5/2}\right].
\end{equation}
To test this expression experimentally, we used the measured data points of $V(t)$ and $z_{\infty}(t)$ shown in Fig.~\ref{fig:filling_profile}(b). The time derivative of the volume appearing on the left-hand side of Eq.~\eqref{Eq:dVdt} is computed using a central-difference scheme, and the equation is tested in Fig.~\ref{fig:Volume_and_Flux} by plotting $\mathrm{d}V/\mathrm{d}t$ as a function of the right-hand side of the equation, omitting the prefactor~$k_e$.
Although this procedure results in large error bars for some data points, the overall trend is consistent with a proportionality relation, yielding a coefficient $k_e = (1.8 \pm 0.7) \times 10^{-2}$. This third method to determine the exchange coefficient~$k_e$, here during the transient stage, leads to a measured value of the same order as the two previous methods discussed in this article, performed in the steady regime.

\subsubsection{Two estimates of the transition time~$t^\star$}

We finally focus on estimating the transition time~$t^\star$, shown experimentally in Fig.~\ref{fig:filling_time}. Two main time scales can be derived from the flux balance equation~\eqref{Eq:bilan_temporal}, depending on the relative strengths of the two fluxes on the right-hand-side. The first time scale corresponds to a filling time~$t_f$, defined as the time needed to fill the meniscus volume through liquid exchange with the soap film. From Eq.~\eqref{Eq:bilan_temporal}, this leads to
\begin{equation}
    t_f \sim \frac{\int_0^H A^\star(z) \textrm{dz} }{\int_0^H q^\star(z)\textrm{dz}}\approx \frac{V(t^\star)}{\mathrm{d}V/\mathrm{d}t},\label{Eq:filling_time_small}
\end{equation}
where the superscript symbol $^\star$ denotes that the quantities are evaluated in the steady state, where they depend only on~$z$. The integral of the section~$A^\star$ along the plate almost represents the half-volume~$V(t^\star)$ of the plate, ignoring the half-volume of the droplet below $z=0$. Indeed, by comparing with numerical simulations, the contribution of the droplet volume to the total volume remains limited, below $20\%$. The time scale~$t_f$ therefore approximately represents the half-volume in the steady regime ($V(t^\star)$) divided by the mean flux ($\mathrm{d}V/\mathrm{d}t$, Eq.~\eqref{Eq:exchange_flux}). For short plates, this approximation seems reasonable since the evolution of $V(t)$ with time is nearly linear, as shown in Fig.~\ref{fig:filling_profile}(b). However, Eq.~\eqref{Eq:filling_time_small} can be tested for the full range of plate heights $H$ by numerically computing  the profile~$r^\star(z)$ through Eq.~\eqref{Eq:main}. 
The predictions of Eq.~\eqref{Eq:filling_time_small} 
are shown in Fig.~\ref{fig:filling_time}. The agreement with the trend of the experimental points is relatively good, showing a rapid decay at small~$H$ followed by a saturation for large~$H$. The agreement is slightly better for short plates, while a factor $2$ is visible between the prediction and the experimental data for long plates.

A second time scale~$t_a$ can be identified from the flux balance equation~\eqref{Eq:bilan_temporal}, by considering the advection of liquid along the meniscus toward the droplet at the bottom of the plate. This advective time scale can be estimated as
\begin{equation}
t_a \sim \int_0^H \frac{A^\star(z)}{Q^\star(z)}\mathrm{d}z = \int_0^H \frac{\mathrm{d}z}{\bar{v}^\star(z)} ,
\label{Eq:advection}
\end{equation}
and represents the time required for a fluid parcel entering the meniscus at the top of the plate to reach the bottom. Again, for simplicity, we consider that the fluid velocity within the meniscus is given by the average velocity profile~$\bar{v}^\star(z)$ in the steady state, as given in Eq.~\eqref{Eq:velocity}. Using the typical meniscus profiles obtained by solving Eq.~\eqref{Eq:main} numerically, $t_a$ is plotted as a function of $H$ in Fig.~\ref{fig:filling_time}, with a blue line. Again, the agreement with the experimental trend is good, with an initial decay for short plates and a saturation for long plates. Equation~\eqref{Eq:advection} leads to an advection time scale~$t_a$ very similar to the filling time scale~$t_f$, given by Eq.~\eqref{Eq:filling_time_small}. 

The time when the meniscus reaches its steady state~$t^\star$ is therefore connected to the filling and advection time scales, defined in Eqs.~\eqref{Eq:filling_time_small} and \eqref{Eq:advection}. These two time scales appears to be very similar for all plate heights, showing that they are both relevant to the way in which the steady state is reached. Indeed, the filling time~$t_f$ somehow takes into account advection of the fluid through the two integrals over the plate height. These two time scales both start to saturate for $H\approx40$~mm, which is of the order of the critical height $H_c$ for the experimental parameters used in Figs.~\ref{fig:filling_profile} and~\ref{fig:filling_time}. This value, that marks the critical height at which steady-state profiles begin to deviate from the hydrostatic shape, therefore also corresponds to the cross-over between the decay of the transition time and its saturation.

\section{Discussion and conclusion}

In this study, we investigate the liquid exchange between vertical menisci and vertical soap films by analysing the shape dynamics of the contact meniscus formed between a solid plate and the film, both in the transient regime and at steady state.

As in the case of rings inserted into vertical soap films \citep{vigna-brummer_flowing_2025}, we observe two distinct regimes: a hydrostatic regime for plates shorter than a critical height $H_c$, and a flowing regime for longer plates, where regions located above the same critical height experience a significant perturbation of the meniscus shape due to liquid influx from the film. This transition is readily identified by examining the meniscus radius of curvature at the top of the plate in steady state, which follows different scaling laws depending on the regime. We propose a model that captures both the steady-state meniscus shape and its transient evolution. For the latter, Surface Evolver simulations were performed to compute the liquid volume contained within the meniscus.

We performed three independent measurements of the flux coefficient $k_e$, two in steady state and one in the transient regime. Within experimental uncertainty, this coefficient is constant and independent of plate geometry, film thickness, and meniscus radius of curvature. Further, we observe no dependence on the inclination of the plate from the vertical, down to $60^\circ$. A weighted average over the three values yields $k_e=(2.4\pm0.3)\times1 0^{-2}$, where the weights are taken as the inverse square of the error bars. 
This value can be compared with the theoretical approach of \citet{gros2021marginal}, presented in the Appendix for the case of identical and regularly distributed TFEs. As shown in Fig.~\ref{Fig:patch_and_ke}(b), our measured value intersects the theoretical curve, confirming that the order of magnitude of the flux coefficient remains the same for vertical menisci. 
If one tentatively extends their framework to the complex and intermittent flows observed at the contact between vertical films and menisci, it would suggest that the fraction $\xi$ of the meniscus–film interface covered by TFEs is either around 0.6 or very close to 1. While the latter value could coincide with the dense pattern observed in experiments (see Fig.~\ref{Fig:setup_profile}(b)), such an extrapolation should be treated with caution, given the significant differences between the idealized configuration considered in the model and the strongly heterogeneous dynamics observed here. 
Notably, at very early times after contact (Fig.~\ref{Fig:first_contact} and Appendix), on a timescale of about 50 ms—much shorter than those investigated in this study—the TFE pattern closely resembles that observed in horizontal films, where TFEs nucleate, grow, and merge.

Both in experiments and Surface Evolver simulations, we observe that the meniscus typically extends below the plate by about one millimeter in steady state. In the model, this extension is accounted for by the position $z_\infty$. As long as the plate height is large compared with $\lvert z_\infty \rvert$, this lower extension does not significantly affect the meniscus shape predicted by the model, except in the region between $z_\infty$ and 0. In this lower region, our two-dimensional model cannot capture the fully three-dimensional accumulation of liquid forming a drop-like structure at the bottom edge of the plate. 
This drop shape, as well as its detachment criterion, could be investigated in a future study, since it corresponds to a pendant-drop geometry subject to specific boundary conditions. We expect its maximal extension to scale with the capillary length $\lambda_c$—consistent with the comparable orders of magnitude observed—but it may also depend on the plate width $W$ when this parameter becomes comparable to or larger than $\lambda_c$. This problem may be relevant to drop condensation in complex geometries, where confinement and edge effects influence drop growth and detachment \citep{lee_water_2012, medici_edge_2014, park_condensation_2016, benilov_capillary_2022, illahie_fog_2025}.

\begin{figure*}
    \centering
    \includegraphics[width=0.5\linewidth]{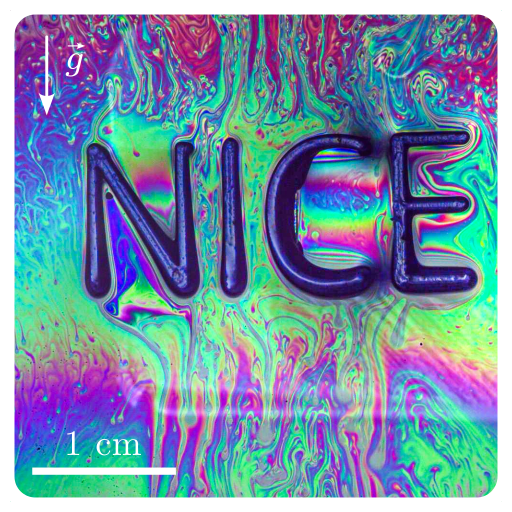}
    \caption{Snapshot of the contact of 3D printed objects, here letters, with a soap film. The corresponding movie is available in the Supplementary Material (movie 6).}
    \label{Fig:Nice}
\end{figure*}

While this study elucidates the meniscus shape in the simple configuration of a plate inserted into a film, the presence of the meniscus also alters the soap film itself. In addition to triggering liquid exchange and film thinning, liquid leakage at the bottom of the plate and the upward migration of thin film elements above it significantly modify the local dynamics. This is illustrated in Fig.~\ref{Fig:Nice} for more complex geometries, where the thickness profile is strongly perturbed by the presence of objects in contact with the film. Such coupling effects are expected to be relevant when solid particles are embedded in soap films or surface bubbles \citep{velikov_direct_1998, shi_spontaneous_2024, louyer_sliding_2025}, or when soap films interact with or are pierced by fibres \citep{whyte_interactions_2017, hjelt_foam_2022, chammouma_crystalline_2025}, with potential implications for the stability of liquid foams \citep{cohen-addad_rigidity_2007, fameau_effect_2014, timounay_gas_2017}.

Finally, quantifying the liquid exchange is essential for a comprehensive description of thinning and drainage in vertical films due to exchange with the lateral menisci. In our experiments, the film was continuously fed with liquid in order to maintain controlled and reproducible conditions. A remaining difficulty is that the film thickness $h$ is not strictly uniform along the meniscus, an effect that is not incorporated into the present model. In practice, the thickness profile is also non-uniform in the fed film, and we measured $h$ at the top of the plate. At leading order, this approximation does not appear to affect our results significantly. Quantitatively, $h$ varies by less than 50\% between $z = H_c$ and $z = H$, i.e. in the region where the influx from the film influences the meniscus profile. Although this variation plays a secondary role in our configuration, it could have a more pronounced impact in freely draining vertical films and may contribute to explaining the emergence of universal thickness profiles \citep{monier_self-similar_2024}. 

\backsection[Supplementary movies]{Supplementary movies are available at ... and described in Appendix~\ref{supp_movies}.}

\backsection[Acknowledgements]{The authors are grateful to Antoine Monier, Isabelle Cantat and Emmanuelle Rio for fruitful discussions.}

\backsection[Funding]{This work was supported by Agence Nationale de la Recherche (ANR-20-CE30-0019). S. J. Cox was partially supported by the Horizon 2020 Framework Programme for Research and Innovation, Grant Agreement number 101008140 EffectFact.}

\backsection[Declaration of interests]{The authors report no conflict of interest.}

\backsection[Author ORCIDs]{ 
A. Vigna-Brummer, https://orcid.org/0009-0005-9470-7975;
S. Cox, https://orcid.org/0000-0001-6129-3394;
M. Argentina, https://orcid.org/0000-0002-8926-7398; 
C. Brouzet, https://orcid.org/0000-0003-3131-3942; 
C. Raufaste, https://orcid.org/0000-0003-4328-7438}

\bibliographystyle{jfm}
\bibliography{Biblio_universalprofile}

\appendix

\section{Exchange coefficient for a horizontal film}

\begin{figure*}
    \centering
    \includegraphics[height=0.33\linewidth]{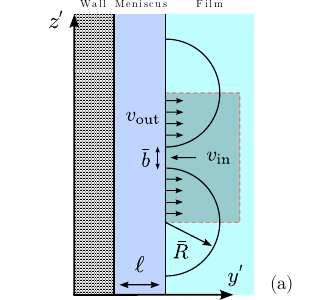}
    \includegraphics[height=0.33\linewidth]{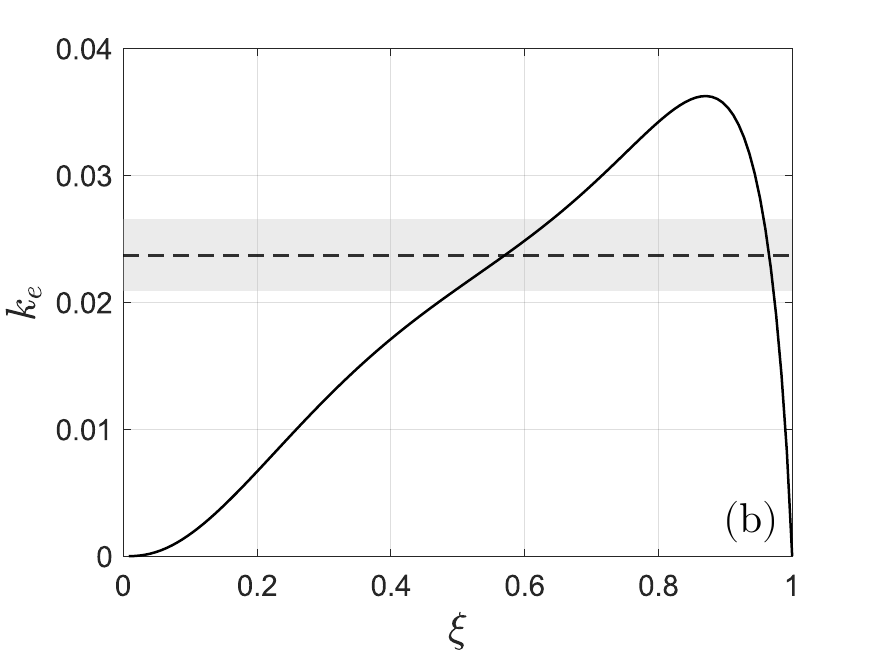}
    \caption{
(a)~Sketch illustrating the growth of thin film elements (TFEs), represented as half-circles, at the contact between a soap film and a meniscus wetting a solid wall. An elementary mesh is represented by a gray area outlined in red.
(b)~Flux coefficient $k_e$ as a function of the coverage fraction $\xi$. The solid black curve represents the theoretical prediction from Eq.~\eqref{Eq:k_e_theorie}. The horizontal dashed line indicates the weighted mean experimental value obtained in this study, $k_e=(2.4\pm0.3)\times10^{-2}$ (from three different methods).
    }
    \label{Fig:patch_and_ke}
\end{figure*} 

The mechanism of marginal regeneration was first modeled by \citet{mysels1959soap} to explain film drainage through exchanges between the film and the meniscus. This concept was recently extended by \citet{gros2021marginal} in a horizontal film, where the TFEs in contact with the meniscus are not subjected to buoyancy forces. 
In this model, thin film elements of thickness~$h_{\mathrm{TFE}}$ are modeled as identical half-circles of average radius~$\bar{R}$, regularly spaced by an average distance~$\bar{b}$, as illustrated in Fig.~\ref{Fig:patch_and_ke}(a). These elements grow within a soap film of thickness~$h > h_{\mathrm{TFE}}$. The quantity $\xi = 2\bar{R}/(2\bar{R} + \bar{b})$ represents the fraction of the meniscus length in contact with thin film elements. 

The exchange of liquid along the film–meniscus contact consists of two complementary processes: portions of the film of thickness~$h$ enter the meniscus at a velocity~$v_{\mathrm{in}}$ through the inter-distance~$\bar{b}$, while thinner films of thickness~$h_{\mathrm{TFE}}$ are extracted from the meniscus within existing TFEs at a velocity~$v_{\mathrm{out}}$. Conservation of film surface then leads to the relation
\begin{equation}\label{Eq:surface_conservation}
   \left(1-\xi\right)v_\text{in}= \xi\,v_\text{out} .
\end{equation}
The liquid exchange flux~$q$, per unit length of the meniscus, from the soap film to the meniscus arises from the difference in thickness between the incoming thicker film and the thinner film extracted from the meniscus, and can be written as
\begin{equation}\label{Eq:linear_flux_Gros2021}
    q=\left(1-\xi\right)v_\text{in}h- \xi\,v_\text{out} h_\text{TFE}=\xi\,v_\text{out}\left(h-h_\text{TFE}\right).
\end{equation}
In this framework, the expressions for $v_{\mathrm{out}}$ and $h_{\mathrm{TFE}}$ depend on $h$ and~$\xi$ and are given by solving the lubrication equation~\citep{gros2021marginal} leading to
\begin{equation}\label{Eq:v_out_et_h_out}
    v_{\text{out}} = \frac{\gamma}{3\eta}f(\xi)\left(\frac{\lambda^3 h}{2r} \right)^{3/2} \ \ \ \mathrm{and} \ \ \ \ h_\text{TFE}=0.64\lambda^3 h f(\xi)^{2/3},
\end{equation}
\begin{equation*}
     \mathrm{with} \ \ \ f(\xi) = \frac{1-\xi}{\xi}\left[\frac{1.85}{2.5}\left(\frac{1-\xi}{\xi}\right)^{2/3}+1.05 \right]^{-9/2} ,
\end{equation*}
where $\gamma$ is the surface tension, $\eta$ the dynamic viscosity, and $r$ the meniscus radius of curvature. The parameter~$\lambda$ was introduced empirically in the model of \citet{gros2021marginal}, with a value close to~$1.8$ providing good agreement with their experimental data.
Substituting the expressions for $v_{\mathrm{out}}$ and $h_{\mathrm{TFE}}$ from Eq.~\eqref{Eq:v_out_et_h_out} into Eq.~\eqref{Eq:linear_flux_Gros2021} recovers the scaling of Eq.~\eqref{Eq:linear_flux}, with a prefactor
\begin{equation}\label{Eq:k_e_theorie}
    k_e = \frac{\xi\,f(\xi)}{3}\left(\frac{\lambda^3}{2}\right)^{3/2}\left(1-0.64\lambda^3f(\xi)^{2/3}\right).
\end{equation}
Remarkably, this coefficient depends solely on the parameter~$\xi$~\citep{gros2021marginal}. 

Figure~\ref{Fig:patch_and_ke}(b) shows the evolution of the flux coefficient $k_e$ as a function of~$\xi$. This figure does not appear in~\citet{gros2021marginal} but was recently presented in the review by~\citet{cantat_drainage_2026}, using slightly different notational conventions. The coefficient $k_e$ vanishes in the two limiting cases $\xi = 0$ and $\xi = 1$, and reaches a maximum value of about~0.036 for $\xi \simeq 0.87$.

\section{TFE nucleation and growth: determination of $\lambda$ at the early instants of contact}

Figure~\ref{Fig:first_contact}(a) presents a high-speed image sequence capturing the initial instants of contact between a vertical glass plate and a soap film. The temporal evolution of the mean TFE area, $\langle A_{\mathrm{TFE}}\rangle$, was quantified through image processing. First, the images were binarized and the TFE contours were detected to determine the total number of elements. The total TFE area was then measured by filling these regions and subtracting the area occupied by the plate width $W$ and the lateral menisci $2\ell$. The mean area $\langle A_{\mathrm{TFE}}\rangle$ was finally obtained by dividing the total area by the number of detected TFEs.

\begin{figure*}
    \centering
    \includegraphics[height=0.33\linewidth]{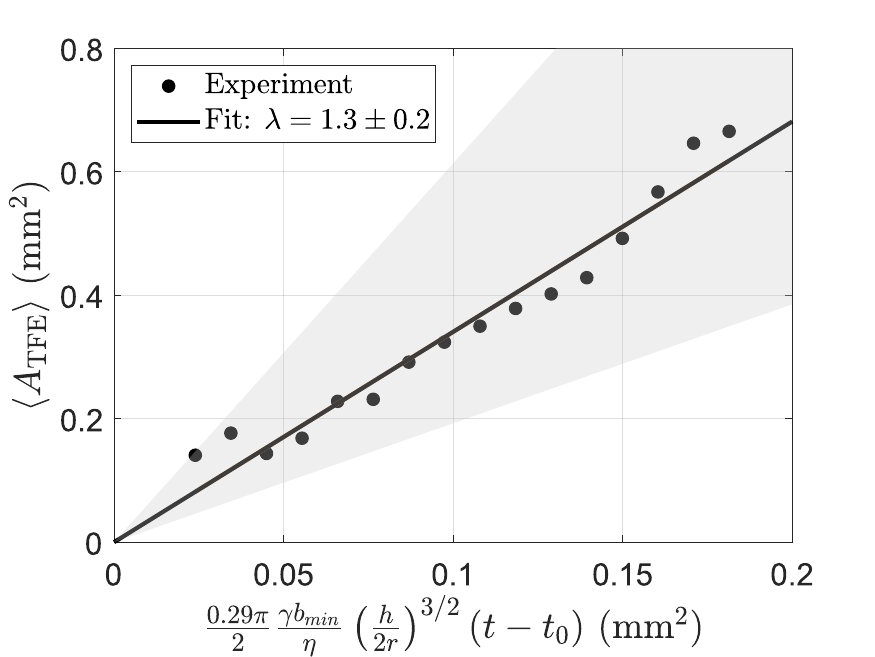}
    \caption{Mean patch area versus the theoretical scaling prediction (Eq.~\ref{Eq:croissane_Patchs}) derived from the image sequence. Black disks represent the results from image analysis. All parameters are fixed by the experimental conditions: $b_\text{min} = (32.4\pm8.1)~\mu$m, $h=(5.5\pm0.2)~\mu$m and $r=(45.6\pm8.1)~\mu$m. The black line is a linear fit with a slope of $\lambda^{9/2}=3.4$. The grey shaded area accounts for the error bounds derived from the parameter uncertainties.}
       \label{Fig:app_first_contact}
\end{figure*}

Figure~\ref{Fig:app_first_contact} shows the results of this analysis. The experimental measurements (black dots) exhibit a clear linear trend, with two notable exceptions. First, at the very beginning of the sequence, the data points lie above the trend line. In this regime, the TFEs are too small to be accurately resolved, leading to an underestimation of their total number. Second, the final data points deviate from the linear growth due to coalescence events (observed between 36 and 39~ms in the image sequence (Fig.~\ref{Fig:first_contact})). These events cause a ‘step-like’ pattern in the evolution of the mean area, as the merging of TFEs locally disrupts the constant growth rate. This effect is more pronounced when there are few TFEs covering large areas. An initial linear fit was used to define the contact time~$t_0 = 4.1$~ms, which is consistent with the image sequence. 

To describe this growth, \citet{gros2021marginal} proposed the following theoretical prediction. The mean TFE area is defined as $\langle A_{\text{TFE}}\rangle = \pi \bar{R}^2/2$. In the absence of coalescence, the temporal evolution of the mean area is given by $\text{d}\langle A_{\text{TFE}}\rangle/\text{d}t = 2\bar{R}v_{\text{out}}$, leading to $(\text{d}\bar{R}/\text{d}t)_{\text{grow}}= 2v_\text{out}/\pi$. When two TFEs of radius $\bar{R}$ coalesce, they form a single TFE of radius $\sqrt{2}\bar{R}$. Consequently, each coalescence event induces a jump in the average radius $\bar{R} \rightarrow \bar{R} + \delta R^c$, where $\delta R^c = (\sqrt{2}-1)\bar{R}/N$ and $N$ represents the TFE number of homogeneous radius~$\bar{R}$. 
During the growth of the TFEs, the mean free space $\bar{b}$ between elements decreases until it reaches a critical value $b_{\text{min}}$, which corresponds to the minimum spacing prior to coalescence. At this threshold, a coalescence event occurs and $\bar{b}$ shifts abruptly as $b_{\text{min}} \rightarrow b_{\text{min}} + \delta b^c$, where $\delta b^c = 2(2-\sqrt{2})\bar{R}/N$.

Assuming a homogeneous distribution of free space between TFEs, the characteristic time between two coalescence events, $\delta t^c$, corresponds to the growth time required for the free space to reach the critical value $b_{\text{min}}$, namely $\delta t^c = \delta b^c / (4v_{\text{out}}/\pi)$. By summing the contributions of steady growth and discrete coalescence events, the total rate of change $\text{d}\bar{R}/\text{d}t = (\text{d}\bar{R}/\text{d}t)_{\text{grow}} + \delta R^c/\delta t^c$ leads to
\begin{equation}\label{Eq:dR/dt_TFE}
    \frac{\text{d}\bar{R}}{\text{d}t} = \frac{2v_{\text{out}}}{\pi(2-\sqrt{2})} \approx 1.09 v_{\text{out}}.
\end{equation}
Experimental observations indicate that the TFEs cover a significant portion of the plate, as  shown in Fig.~\ref{Fig:first_contact}, with the coverage fraction $\xi = 1 - b_{\text{min}}/(2\bar{R})$ approaching unity, consistent with results reported for horizontal films~\citep{gros2021marginal}. Under these conditions, taking $f(\xi) \approx 0.8 \, b_{\text{min}}/\bar{R}$ and integrating Eq.~\eqref{Eq:dR/dt_TFE} yields
\begin{equation}\label{Eq:croissane_Patchs}
\langle A_{\mathrm{TFE}}\rangle(t) = \frac{0.29\pi}{2}\frac{\gamma b_\text{min}}{\eta}\left(\frac{\lambda^3 h}{2r}\right)^{3/2} t + \frac{\pi R_0^2}{2},
\end{equation}
where $R_0$ is the initial radius of the TFEs at $t=0$, and $\lambda$ is a dimensionless prefactor. While the numerical work of \citet{mysels1959soap} suggests $\lambda=1$, experimental studies by \citet{gros2021marginal} reported values between $1.6$ and $1.8$.

In our analysis, all parameters except $\lambda$ were fixed by the experimental conditions: the critical spacing $b_\text{min} = (32.4 \pm 8.1)~\mu$m is on the order of one pixel (corresponding to the spatial resolution limit and accounting for the primary source of uncertainty), the film thickness is $h = (5.5 \pm 0.2)~\mu$m, the radius of curvature $r = (45.6\pm8.1)~\mu$m represents the mean value over the sequence and $R_0=0$. The solid black line in Fig.~\ref{Fig:app_first_contact} represents the best fit to the data, yielding $\lambda = 1.3\pm0.2$. The shaded grey region illustrates the uncertainty propagated from the experimental errors in $h$, $b_\text{min}$, and $r$.

\section{Description of the supplementary movies\label{supp_movies}}

Movie 1 shows a video of a draining film (no liquid supply), with parameters $H = 26$~mm, $W = 1$~mm and $\alpha = 0^\circ$. It  corresponds to the experiment in Fig.~\ref{Fig:setup_profile}(b). 
The movie is shown in real time.

Movie 2 shows a video of a draining film (no liquid supply), with parameters $H = 26$~mm, $W = 1$~mm and $\alpha = 20^\circ$. 
It corresponds to the experiment in Fig.~\ref{Fig:setup_profile}(c). 
The movie is slowed down by a factor of $2$.

Movie 3 shows a video corresponding to the experiment in Fig.~\ref{Fig:first_contact}, with parameters $H = 26$~mm, $W = 150,\mu$m, $\alpha = 0^\circ$ and $h=(5.5\pm0.2)~\mu$m. It captures the very first instants after the plate is introduced into the film. The movie is slowed down by a factor of $333$.

Movie 4 shows a video of the long-time transient regime, with parameters $H = 14.5$~mm, $W = 1$~mm, $\alpha = 0^\circ$ and $h=1.6~\mu$m. It corresponds to the experiment in Fig.~\ref{Fig:meniscus_growth}. The movie is accelerated by a factor of $10$.

Movie 5 shows a video corresponding to a numerical simulation performed with Surface Evolver, with parameters $H = 14.5$~mm, $W = 1$~mm, and $\alpha = 0^\circ$. The volume is progressively increased up to the maximum value prior to the onset of instability. It corresponds to Surface Evolver simulations in Fig.~\ref{fig:filling_profile}(a).

Movie 6 shows a video corresponding to the experiment in Fig.~\ref{Fig:Nice}, with $h\sim1~\mu$m. The movie is shown in real time.

\end{document}